\documentclass[aip,reprint]{revtex4-1}

\usepackage{xcolor}
\usepackage{float}

\usepackage{graphicx}
\usepackage{dcolumn}
\usepackage{bm}

\usepackage{amsmath}

\usepackage[utf8]{inputenc}
\usepackage[T1]{fontenc}
\usepackage{mathptmx}
\usepackage{etoolbox}
\usepackage[normalem]{ulem} 

\usepackage{xr}
\makeatletter
\newcommand*{\addFileDependency}[1]{
  \typeout{(#1)}
  \@addtofilelist{#1}
  \IfFileExists{#1}{}{\typeout{No file #1.}}
}
\makeatother

\newcommand*{\myexternaldocument}[1]{%
    \externaldocument{#1}%
    \addFileDependency{#1.tex}%
    \addFileDependency{#1.aux}%
}

\myexternaldocument{suppl}

\draft 

\begin{document}


\title{Exchange Correlation Potentials from Full Configuration Interaction in a Slater Orbital Basis} 



\author{Soumi Tribedi}
\affiliation{Department of Chemistry, University
of Michigan, Ann Arbor, Michigan 48109, United States}
\affiliation{Michigan Institute for Data Science, University of Michigan, Ann Arbor, Michigan 48109, United States}

\author{Duy-Khoi Dang}
\affiliation{Department of Chemistry, University
of Michigan, Ann Arbor, Michigan 48109, United States}

\author{Bikash Kanungo}
\affiliation{Department of Mechanical Engineering,
University of Michigan, Ann Arbor, Michigan 48109, United
States}

\author{Vikram Gavini}
\affiliation{Department of Mechanical Engineering,
University of Michigan, Ann Arbor, Michigan 48109, United
States}
\affiliation{Department of Materials Science and Engineering, University of Michigan, Ann Arbor, Michigan 48109, United
States}

\author{Paul M. Zimmerman}
\email{paulzim@umich.edu}
\affiliation{Department of Chemistry, University
of Michigan, Ann Arbor, Michigan 48109, United States}


\date{\today}

\begin{abstract}
Ryabinkin-Kohut-Staroverov (RKS) theory builds a bridge between wave function theory and density functional theory by using quantities from the former to produce accurate exchange-correlation potentials needed by the latter. In this work, the RKS method is developed and tested alongside Slater atomic orbital basis functions for the first time. To evaluate this approach, Full Configuration Interaction computations in the Slater orbital basis are employed to give quality input to RKS, allowing full correlation to be present along with correct nuclei cusps and asymptotic decay of the wavefunction. SlaterRKS is shown to be an efficient algorithm to arrive at exchange correlation potentials without unphysical artifacts in moderately-sized basis sets. Furthermore, enforcement of the nuclear cusp conditions will be shown to be vital for the success of the Slater-basis RKS method. Examples of weakly and strongly correlated molecular systems will demonstrate the main features of SlaterRKS. 
\end{abstract}


\maketitle 

\section{Introduction}
Density Functional Theory (DFT)\cite{mardirossian_thirty_2017,becke_perspective_2014,burke_dft_2013}, particularly within the Kohn Sham (KS) formalism,\cite{PhysRev.136.B864} provides a cost-effective, scalable means for approximating the quantum behavior of electronic states. The KS equations are

\begin{equation}
    \bigg[-\frac{1}{2}\nabla^2 + v(\mathbf{r}) + v_H(\mathbf{r}) + v_{XC}(\mathbf{r})\bigg]\phi_i(\mathbf{r}) = \epsilon_i \phi_i(\mathbf{r}),
    \label{kseq}
\end{equation}

and $\phi_i$ are the KS orbitals, which together comprise a system of non-interacting electrons having the  density $\rho^{KS}(\mathbf{r}) = \sum_i n_i |\phi_i(\mathbf{r})|^2$.\cite{PhysRev.140.A1133} The orbital energies, $\epsilon_i$, arise from the kinetic energy and three potentials: $v(\mathbf{r})$ is the external potential (usually the nuclear potential), $v_H(\mathbf{r})$ is the Hartree potential which accounts for the classical Coulomb interaction between electrons, and $v_{XC}(\mathbf{r})$ is the exchange-correlation potential which contains the non-classical contribution from the kinetic energy and electron-electron repulsion. The unknown component of DFT, $v_{XC}(\mathbf{r})$, is defined as the functional derivative of the XC energy functional,
\begin{equation}
    v_{XC}(\mathbf{r}) = \frac{\delta E_{XC}[\rho(\mathbf{r})]}{\delta \rho(\mathbf{r})}.
    \label{excderiv}
\end{equation}
The functional of the density, $E_{XC}[\rho(\mathbf{r})]$, and its derivative in Eq. \ref{excderiv} are approximated in practice. Together, Eqs. \ref{kseq} and \ref{excderiv} are used to solve for KS orbitals, which provide a description of the electron density and its corresponding energy.

The choice of the approximate form for $E_{XC}[\rho(\mathbf{r})]$ is made from the various rungs of the 'Jacob's Ladder' of functionals that have been developed over the past half century.\cite{doi:10.1063/1.1904565} The capabilities of DFT for molecules \cite{cohen_assessment_2000,kummel_orbital-dependent_2008,sun_accurate_2016} and solids\cite{distasio_individual_2014,marsman_hybrid_2008,isaacs_performance_2018} have undoubtedly increased over this time, but  questions still remain regarding well-documented problems with self-interaction, delocalization errors, and strong correlation. \cite{cohen_challenges_2012,verma_status_2020,crisostomo_seven_2022,bryenton_delocalization_nodate} Since these errors are tied closely to the electron density, it is clear that improvements to $v_{XC}(\mathbf{r})$ are needed to improve upon current DFT approximations.

The focus of this article is therefore on the exchange-correlation potential, since knowledge about this potential could lead to improved functional approximations. \cite{green_learnings_1998,tozer_development_1998,wilson_hybrid_2001,menconi_emphasizing_2001,gaiduk_self-interaction_2012} This point has been emphasized by recent work showing that widely used functionals give rise to potentials that are far from the exact (even for SCAN0 which has $v_{XC}$ closest to the exact, the errors are in the range of $O(10^{-1}-10^0)$).\cite{kanungo_comparison_2021} These errors may be attributed to the standard practice of functional training to reproduce energies, while neglecting errors in the exchange-correlation potential. Providing this information, however, requires means to accurately compute $v_{XC}(\mathbf{r})$, which is a highly nontrivial task.

For a given reference density, the corresponding XC potential can in principle be obtained numerically by the inverse DFT approach.\cite{shi_inverse_2021} The inverse problem maps a given density to its corresponding potential through a unique one-to-one mapping.\cite{PhysRev.136.B864,runge_density-functional_1984} The inverse relation, however, is well posed only in a complete basis.\cite{kohn_v-representability_1983,hadamard_1902} In an incomplete basis the unique mapping of potential to density does not hold true, i.e., different XC potentials can map to the same density.\cite{harriman_densities_1986,harriman_density_1990,staroverov_optimized_2006} In practice, inverse calculations are often performed using incomplete basis sets (such as finite Gaussian basis sets) and persistent numerical problems result. Recently, Kanungo et al. implemented a complete, finite-element basis and used it to obtain highly accurate potentials,\cite{kanungo_exact_2019} and St\"{u}ckrath et al. alleviated the problem using a multiresolution wavelet basis to similar effect.\cite{stuckrath_reduction_2021} These recent results show the possibilities of complete basis sets in tackling the inverse DFT problem, but have not improved upon the situation for finite basis sets.

Elegantly skipping past the inverse DFT problem, Staroverov and coworkers developed means to obtain $v_{XC}(\mathbf{r})$ from the reduced density matrices (RDMs) of any wavefunction method.\cite{ryabinkin_reduction_2015,cuevas-saavedra_kohnsham_2015} The RKS method evaluates $v_{XC}(\mathbf{r})$ from the comparison of two local energy balance equations, one originating from the KS equations and the other from wavefunction theory. Since the RKS method uses the wavefunction instead of the density to obtain the potential and employs systematic approximations to the kinetic energy density, it is less prone to finite-basis-set errors as compared to related techniques. \cite{kumar_accurate_2020, shi_n2v_2022} Using RKS, potentials can be straightforwardly derived for atoms and molecules in moderate or larger Gaussian basis sets. Even though densities expanded in Gaussian type orbitals (GTOs) have inherent shortcomings, the stability of the RKS (and mRKS) methods in obtaining  accurate exchange-correlation potentials is a promising step forward for finite-basis DFT. To go further with the RKS method, we envisioned improving the wave function, electron densities, and $v_{XC}(\mathbf{r})$ by the use of Slater type orbitals (STO). \cite{schipper_kohn-sham_1997}

STOs have good physical properties that are expected for electronic states, for instance the ability to describe cusps at the nuclei and at long range, exponentially decaying tails. \cite{kato_eigenfunctions_1957,helgaker_molecular_2000,reinhardt_cusps_2009} Particularly in the construction of densities and the corresponding potentials, these properties make STOs a better choice than GTOs, which suffer from unphysical oscillatory behaviour under the Laplacian.\cite{schipper_kohn-sham_1997} While STOs have seen less use than GTOs in quantum chemistry, their limitations can be traced to the difficult two-electron integral evaluation, which must be done numerically in 6 dimensions. Recently Zimmerman and coworkers developed an efficient GPU-accelerated algorithm to evaluate STO integrals in the resolution-of-the-identity (RI) approximation.\cite{dang_numerical_2022} This opened up the opportunity for us to carry out the RKS procedure with STO basis sets.

By using STO basis sets, one important property of the electron density can be naturally incorporated into the RKS procedure. The Kato cusp condition specifies how the density must behave near the nucleus.\cite{kato_eigenfunctions_1957} This condition is met when every occupied molecular orbital satisfies
\begin{equation}
    \frac{\partial \phi_i}{\partial \mathbf{r}}\bigg|_{\mathbf{r}=R_B} = -Z_B\phi_i(R_B),
    \label{kato}
\end{equation}
where $R_B$ is the position of a nucleus $B$ with atomic number $Z_B$. With STOs, the cusp condition can be enforced by using a modified SCF procedure\cite{handy_molecular_2004} which will be described in the Methods section. As a result, the singularity due to electron-nuclear attraction at the nucleus will be compensated by the kinetic energy. This property appears unreachable in finite GTO basis sets, and its effect on $v_{XC}(\mathbf{r})$ near the nucleus will be an interesting question to examine (Section \ref{result}).

This article presents exchange correlation potentials for atoms and molecules, derived from highly accurate full configuration interaction (FCI) densities using Slater type orbitals. FCI wave functions in STO basis sets capture all dynamic and static correlation available to the basis, with the additional benefit of having the correct nuclear cusp and long range asymptotic behavior. The ability to calculate XC potentials at this high level of theory will therefore be examined for the first time in the results that follow.

\section{Method}

The RKS method utilizes the one particle and two particle reduced density matrices (1-RDM and 2-RDM respectively) of the wavefunction to evaluate $v_{XC}$.\cite{ryabinkin_reduction_2015} The notations $WF$ or $KS$ are used throughout this manuscript to denote the terms derived from wavefunction or KS-DFT methods, respectively. The RKS working equation,
\begin{equation}
    v_{XC}(\mathbf{r}) = v_{XC,Slater}^{WF}(\mathbf{r}) + \frac{\tau^{WF}(\mathbf{r})}{\rho^{WF}(\mathbf{r})} - \frac{\tau^{KS}(\mathbf{r})}{\rho^{KS}(\mathbf{r})} + \epsilon^{KS}(\mathbf{r}) - \epsilon^{WF}(\mathbf{r}),
    \label{rks}
\end{equation}
is derived from local energy balance equations under the condition $\rho^{WF}(\mathbf{r}) = \rho^{KS}(\mathbf{r})$. Here, $\rho^{WF}(\mathbf{r})$ is fixed to that of the wavefunction reference, and $\rho^{KS}(\mathbf{r})$ is the density from the self-consistent solutions of the KS equations (Eq. \ref{kseq}). The Slater exchange-correlation charge potential, $v_{XC,Slater}^{WF}(\mathbf{r})$,\cite{slater_generalized_1953} is expressed as,
\begin{equation}
    v_{XC,Slater}^{WF}(\mathbf{r}) = \int \frac{\rho^{WF}_{XC}(\mathbf{r},\mathbf{r}_2)}{|\mathbf{r}-\mathbf{r}_2|}d\mathbf{r}_2,
    \label{vxch}
\end{equation}
where, $\rho^{WF}_{XC}(\mathbf{r})$ is the exchange-correlation hole density derived from the relation, $\Gamma(\mathbf{r},\mathbf{r}_2;\mathbf{r},\mathbf{r}_2) = P(\mathbf{r},\mathbf{r}_2) = \frac{1}{2} \rho^{WF}(\mathbf{r})\big[\rho^{WF}(\mathbf{r}_2) + \rho^{WF}_{XC}(\mathbf{r},\mathbf{r}_2)\big]$. Here $\Gamma(\mathbf{r},\mathbf{r}_2;\mathbf{r},\mathbf{r}_2)$ is the special case of the coordinate representation of the 2-RDM, 
\begin{equation}
\Gamma(\mathbf{r},\mathbf{r}_2;\mathbf{r}^\prime,\mathbf{r}_2^\prime)|_{\mathbf{r} = \mathbf{r}^\prime; \mathbf{r}_2 = \mathbf{r}_2^\prime} = \sum_{pqrs} \Gamma_{pqrs}\psi_p(\mathbf{r})\psi_q(\mathbf{r}_2)\psi_r^*(\mathbf{r}^\prime)\psi_s^*(\mathbf{r}_2^\prime),
\end{equation}
where $\Gamma_{pqrs}$ are the matrix elements of the orbital representation of the 2-RDM.
The positive-definite kinetic energy densities, $\tau(\mathbf{r})$, and the average local electron energies, $\epsilon(\mathbf{r})$, are defined as, 

\begin{equation}
    \tau^{WF}(\mathbf{r}) =\frac{1}{2}\sum_in_i|\nabla\psi_i(\mathbf{r})|^2,
    \label{tauwf}
\end{equation}
\begin{equation}
    \tau^{KS}(\mathbf{r}) = \frac{1}{2}\sum_i n_i|\nabla\phi_i(\mathbf{r})|^2,
    \label{p}
\end{equation}
\begin{equation}
    \epsilon^{WF}(\mathbf{r}) = \frac{1}{\rho^{WF}(\mathbf{r})}\sum_{j} \lambda_{j}|\psi_j(\mathbf{r})|^2,
    \label{epsilonwf}
\end{equation}
\begin{equation}
    \epsilon^{KS}(\mathbf{r}) = \frac{2}{\rho^{KS}(\mathbf{r})}\sum_{i=1}^{N/2} \epsilon_i|\phi_i(\mathbf{r})|^2,
    \label{epsilonks}
\end{equation}
where the orbitals $\psi_j$ are associated with generalized Fock eigenvalues $\lambda_{j}$, and $\phi_i$ are the set of KS orbitals with eigenvalues $\epsilon_i$. Other choices for the kinetic energy densities are the Laplacian and Pauli forms. In the original RKS method, the positive-definite kinetic energy was used rather than the Laplacian kinetic energy ($\tau_L^{KS} = -\frac{1}{2}\sum_i n_i \phi_i^*(\mathbf{r})\nabla^2\phi_i(\mathbf{r})$) which are related by $\tau_L = \tau -0.25 \nabla^2 \rho$.\cite{ryabinkin_reduction_2015,cuevas-saavedra_kohnsham_2015} This choice can be motivated by equating KS and WF densities (i.e. $\rho^{KS} = \rho^{WF}$) or by observing that the positive-definite kinetic energy is more numerically stable near an atom. In the modified RKS method, the Pauli kinetic energy density ($\tau_P = \tau - |\nabla\rho|^2/8\rho$)\cite{ospadov_improved_2017} is used to proceed further along this path. For Slater-based RKS, two forms of kinetic energy densities were evaluated (Laplacian and positive definite), and it was found that the positive definite form has better properties (vide infra).

The above RKS equations give $v_{XC}(\mathbf{r})$ up to an arbitrary constant. At the asymptotic limit of $\mathbf{r}\to\infty$, $v_{XC}(\mathbf{r}) \to v_{XC,Slater}^{WF}(\mathbf{r}) \sim -1/\mathbf{r}$. $\epsilon^{KS}$ and $-\tau^{KS}/\rho^{KS}$ approach $\epsilon_{HOMO}$, whereas the analogous wavefunction terms approach $-I_{min}$, the first ionization energy from extended Koopman's theorem. In order to enforce the asymptotic decay of $v_{XC}(\mathbf{r})$ (and determine the constant), all the terms to the right of Eq. \ref{rks} except for the $v_{XC,Slater}^{WF}$ must cancel at $\mathbf{r}\to\infty$. Therefore all the eigenvalues of the KS orbitals are shifted such that $\epsilon_{HOMO} = -I_{min}$ is satisfied. Further, in the far field, where the densities are almost zero, the numerical evaluation of the kinetic energy densities, i.e. the second and third terms in Eq. \ref{rks} yields unphysically large values. Since at large distances $v_{XC}$ and $v_{XC,Slater}^{WF}$ are expected to be identical, a function ($F$) is used to smoothly transition $v_{XC}$ into $v_{XC,Slater}^{WF}$ at the threshold of low density ($\theta = 10^{-5}$),
\begin{equation}
    v_{XC}(\mathbf{r})^{smooth} = F(\mathbf{r}) v_{XC}(\mathbf{r}) + (1-F(\mathbf{r}))v_{XC,Slater}^{WF}(\mathbf{r})
\end{equation}
where,
\begin{equation}
    F(\mathbf{r}) = \frac{\rho(\mathbf{r})}{\rho(\mathbf{r})+\theta}
    \label{smooth}
\end{equation}

\begin{figure}[h]
\includegraphics[width=\linewidth]{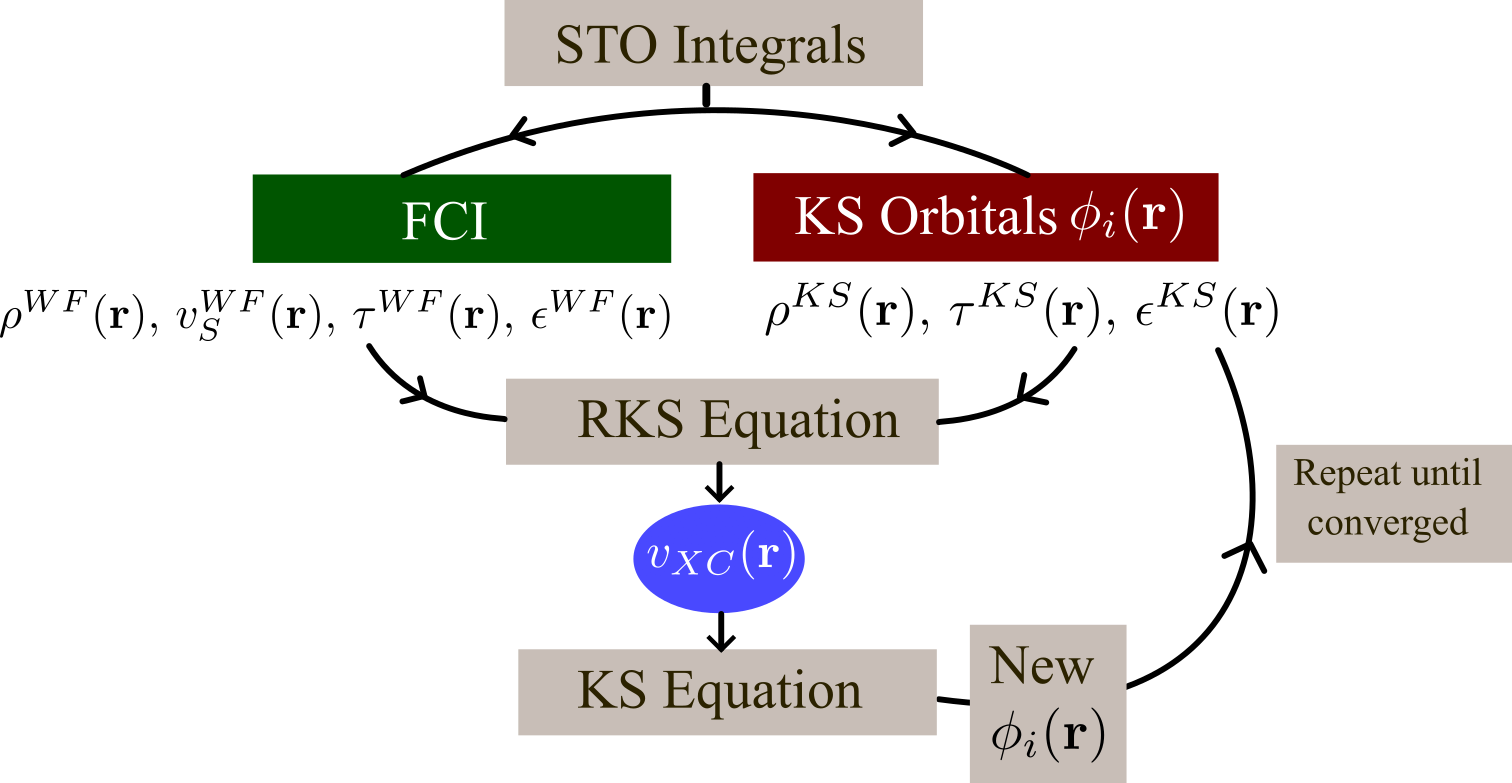}
\caption{\label{fig:alg} The algorithm of the RKS method using STO integrals}
\end{figure}

The algorithm of the RKS method (Fig. \ref{fig:alg}) entails evaluating the wavefunction terms first, i.e., $v_{XC,Slater}^{WF}(\mathbf{r})$, $\rho^{WF}(\mathbf{r})$, $\tau^{WF}(\mathbf{r})$, and $\epsilon^{WF}(\mathbf{r})$. Next, an initial guess for the KS orbitals is chosen and the corresponding terms, $\rho^{KS}(\mathbf{r})$, $\tau^{KS}(\mathbf{r})$ and $\epsilon^{KS}(\mathbf{r})$ are evaluated and the $\epsilon_i$ are shifted with respect to $I_{min}$. Eq. \ref{rks} then provides $v_{XC}(\mathbf{r})$. This potential is then used to solve for new KS orbitals from Eq. \ref{kseq}, and the KS terms are updated in Eq. \ref{rks} until $v_{XC}(\mathbf{r})$ and the KS orbitals become self-consistent.

In all, the SlaterRKS procedure follows the original RKS procedure that used a Gaussian basis set, except for a few details. In particular, it is notable that due to the correct short- and long-range behavior of Slater orbitals, the Laplacian kinetic energy operator, $\tau^{KS}_{L}(\mathbf{r})$ may be a tractable choice. This will be the case only when the conditions of the next paragraph are under consideration.

The molecular orbitals used for the WF and KS theories are constrained to obey Kato's cusp condition (Eq. \ref{kato}) using Handy's method.\cite{handy_molecular_2004} Each molecular orbital is expanded in terms of the basis functions ($\chi$), i.e. $\phi_i = \sum_{\alpha A}c_{\alpha A,i}\chi_{\alpha A}$, where the sum is over exponent $\alpha$ and the atom $A$ on which the basis functions are centered, and $c_i$ are the MO coefficients corresponding to the $i^{th}$ MO. Kato's condition gives
\begin{equation}
    \sum_{\alpha A} p_{\alpha A,B}c_{\alpha A,i} = 0,
    \label{cond1}
\end{equation}
where,
\begin{equation}
    p_{\alpha A,B} = \frac{\partial \chi_{\alpha B}}{\partial \mathbf{r}_B}\bigg|_{\mathbf{r}_B = 0} \delta_{AB} + Z_B\chi_{\alpha A}(\mathbf{r}_B)
\end{equation}
The condition of Eq. \ref{cond1} is enforced by using a modified SCF procedure with,
\begin{equation}
    [(\mathbf{I}-\mathbf{A})(\mathbf{F}-\epsilon\mathbf{S})(\mathbf{I}-\mathbf{A})]\mathbf{c} = 0,
    \label{cuspSCF}
\end{equation}
where $\mathbf{A} = \sum_{BC}\hat{\mathbf{p}}_B (\hat{\mathbf{p}}_B^T \hat{\mathbf{p}}_C)^{-1} \hat{\mathbf{p}}_C^T$. This procedure ensures that all MOs in the correlated wavefunction have correct electron-nuclear cusps, and the same will be true when solving the KS equations. 

\section{Computational Details}
The SlaterRKS method was implemented in C++ and interfaced with the SlaterGPU\cite{dang_numerical_2022} library to evaluate integrals for Slater basis functions using GPU acceleration. All Coulomb integrals will therefore be evaluated using the Resolution of the Identity (RI) approximation.\cite{dunlap_approximations_1979,werner_fast_2003,distasio_jr_improved_2007} The orbitals and RDMs for the reference WFs come from the heat-bath configuration interaction (HBCI) procedure\cite{holmes_heat-bath_2016,sharma_semistochastic_2017,li_fast_2018,dang_advances_2023,chien_excited_2018} using a tight threshold (He, $\mathrm{H_2}$, LiH: $10^{-5}$ Ha  and $\mathrm{H_2O}$, $\mathrm{CH_2}$: $10^{-4}$ Ha) for accuracy of the variational wavefunction. For atoms and small molecules, the RDMs from this approach will be essentially FCI quality. 

The primary working equation in the SlaterRKS method is Eq. \ref{rks}, where all the quantities are evaluated according to their expressions given in Eqs \ref{vxch}-\ref{epsilonks}. Evaluation of these equations is accelerated on GPU using OpenACC to generate their contribution to $v_{XC}$ on the grid. For Eqs \ref{tauwf}-\ref{epsilonks}, a simple \texttt{acc parallel} directive is prepended to each for-loop over the grid points. Evaluation of $v_{XC,Slater}^{WF}(\mathbf{r})$ via Eq. \ref{vxch} is the most expensive step due to the 6-dimensional integration over two electrons. To minimize wall-time, OpenMP and OpenACC are jointly used to compute $v_{XC,Slater}^{WF}(\mathbf{r})$. Therefore, contributions from Eq. \ref{vxch} involve two nested for-loops, each over the entire grid.  The outer for-loop is parallelized with OpenMP where one thread is launched for each GPU. The inner for-loop is then parallelized with OpenACC using the \texttt{acc parallel} directive with a \texttt{reduction} clause. The \texttt{reduction} clause indicates a summation in the inner for-loop, which reflects the numerical evaluation of the integral in Eq. \ref{vxch}. Since the same grid and integral weights are used for each inner for-loop, they only need to be generated once before entering the outer for-loop of Eq. \ref{vxch}.

The Slater basis sets were taken from the set developed by Baerends and coworkers.\cite{van_lenthe_optimized_2003} Four types of basis set are included in this set, namely DZP, TZP, TZ2P and QZ4P. The DZP, TZP and TZ2P basis sets have double zeta-core functions, while the QZ4P has a triple-zeta core. In the valence region, the DZ is double zeta, TZP and TZ2P triple zeta, and QZ4P quadruple zeta, all with valence polarization functions. For example, the QZ4P basis has 3$\times$1s,  4$\times$2s, 4$\times$2p, 2$\times$3d and 2$\times$4f functions for C atom. Larger basis sets were needed to test the cusp condition implementation on the helium atom (vide infra). 5Z6P and 6Z6P basis sets were thus created in an even-tempered manner, following the procedure of Baerends and coworkers\cite{chong_even-tempered_2004} (see Supporting Information for full specification of these basis sets).

The geometries and ionization energies ($I_{min}$) are taken from the CCCBDB database\cite{cccbdb} and provided in the Supporting Information. The value of $I_{min}$ can be set to the ionization energy from simulation or to the experimental value. Herein, $I_{min}$ is set to the experimental value (given in the SI), as the FCI result is expected to be similar.  The STO integral evaluation as well as the RKS procedure is carried out on numerical atom-centered grid.\cite{becke_multicenter_1988,mura_improved_1996,murray_quadrature_1993} The three dimensional grid is composed of products of radial\cite{mura_improved_1996} and angular\cite{lebedev_quadratures_1976} points with weights according to the Becke partitioning scheme.\cite{becke_multicenter_1988} In all the calculations in this manuscript, 50 radial and 5810 angular points are employed for each atom. 

The convergence of each SlaterRKS run is verified by the $L_1$ and $L_2$ errors in the KS density compared to the reference WF. 
The norms, $\Delta X_{L_1}$ and $\Delta X_{L_2}$ of a property $X$ are defined as,
\begin{equation}
    \Delta X_{L_1} = \int \big|X^A(\mathbf{r})-X^B(\mathbf{r})\big| d\mathbf{r}
\end{equation}

\begin{equation}
    \Delta X_{L_2} = \sqrt{\int (X^A(\mathbf{r})-X^B(\mathbf{r}))^2 d\mathbf{r}}
\end{equation}
where $A$ and $B$ are the reference and calculated values. The calculation is deemed converged when the $\Delta \rho_{L_1}$ does not change more than $10^{-5}$ from one iteration to the next.


\section{Results and Discussion}
\label{result}

In this section the SlaterRKS method is applied to a few prototypical test systems along with a few strongly correlated test cases. The simplest example is the two electron case of hydrogen molecule at three separate bond distances--equilibrium, twice the equilibrium and fully dissociated. The next example is the heteronuclear LiH molecule. In addition the water molecule is examined as a small polyatomic, followed by the more challenging multireference test case of singlet methylene ($\mathrm{CH_2}$). For all of these examples, FCI wavefunctions in the Slater orbital basis sets will be used (unless otherwise noted) as the reference for the RKS procedure.

Before delineating these examples, it is emphasized that the enforcement of the long-range asymptotic behavior is important to reach meaningful exchange correlation potentials. The correct asymptotic decay of $v_{XC}$ follows $-1/\mathbf{r}$,\cite{schmidt_one-electron_2014,wu_density-functional_2003}, though LDA and GGA functionals fail to satisfy this condition.\cite{staroverov_tests_2004} Properties such as the energy of the HOMO, which is tied closely to the ionization energy, will only be accurate with correct asymptotics.\cite{kraisler_asymptotic_2020} In the RKS method, this condition is enforced by shifting $v_{XC}$ such that $\epsilon_{HOMO} = -I_{min}$. In SlaterRKS, an additional condition (see method section) ensures a smooth transition  $v_{XC}(\mathbf{r}) \to v_{XC,Slater}^{WF}(\mathbf{r})$ as $r \to \infty$. The Supporting Information shows the $\mathrm{H_2}$(eq.) molecule, where $v_{XC}$ and $v_{XC,Slater}^{WF}$ decay as $-1/\mathbf{r}$ at low density regions far from the nuclei.

While the cusp condition (Eq. \ref{kato}) will be applied for most of the SlaterRKS results of this work, it will first be evaluated for the helium atom in a range of basis sets. Numerical evaluation of Kato's equation on a few grid points near the nucleus is shown in Tables \ref{tab:cusptz}-\ref{tab:cusp6z} in the SI. These results confirm the cusp enforcement reduces the error at $\mathbf{r}=10^{-6}$ (i.e. at the nuclear position) by two orders of magnitude or more compared to densities without the cusp condition. Obtaining this accuracy near the nucleus, however, is not free because one degree of freedom per atom is lost in the basis when the constraint is applied. Therefore the dependence of the SlaterRKS results on the basis set size needs to be examined before applying the cusp condition more widely.

\begin{figure}[H]
    \includegraphics[width=\linewidth]{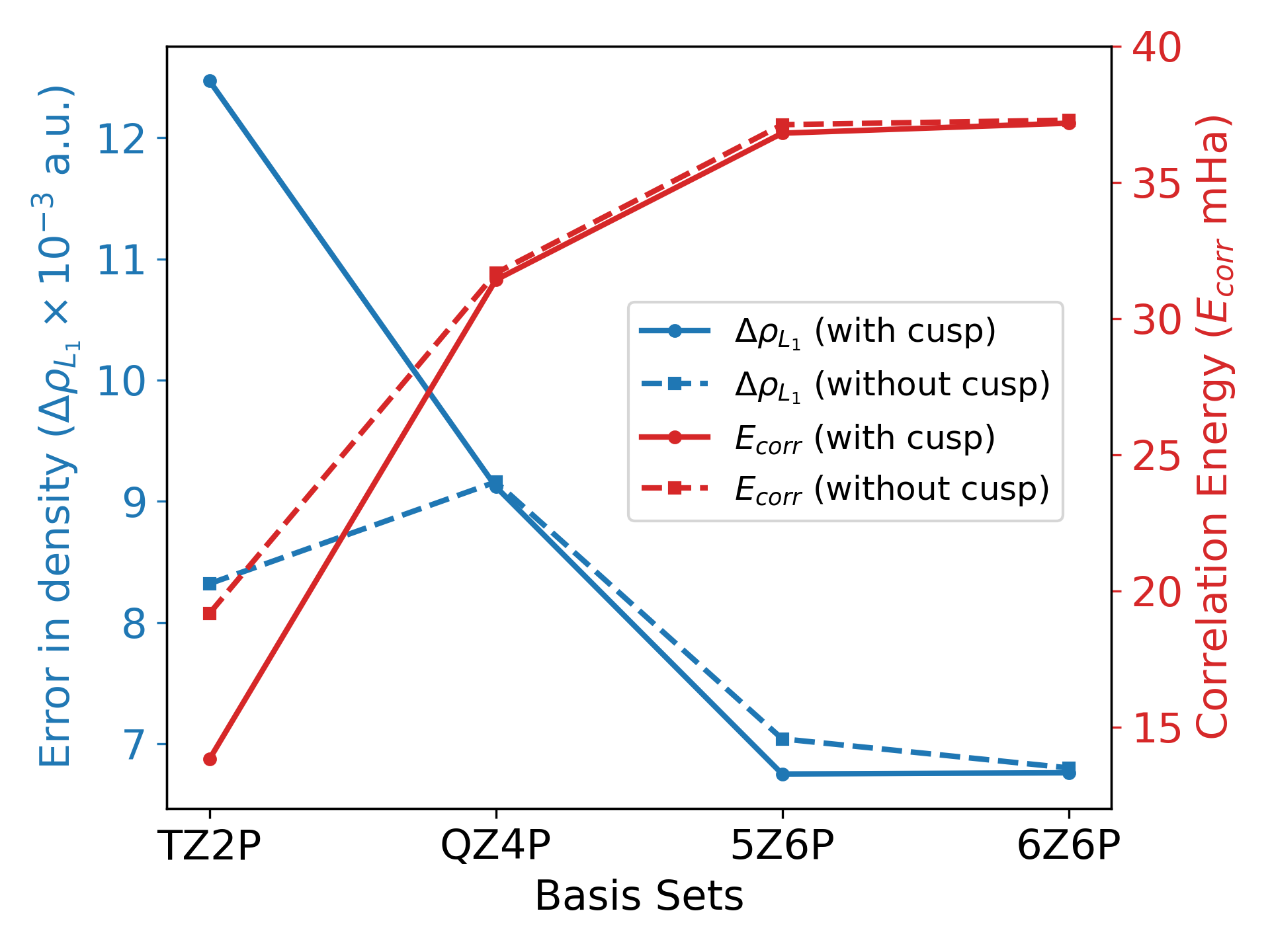}
    \caption{\label{fig:cusp_basis} Results for He atom in various basis sets. Left-hand axis: Comparison of $L_1$ errors of RKS densities with respect to CI densities when the cusp condition is applied vs. not applied. Right-hand axis: Correlation energies.}
\end{figure}

Fig. \ref{fig:cusp_basis} shows the $L_1$ error of density from the SlaterRKS procedure (blue) and the corresponding FCI correlation energy ($E_{corr}$) (red) with cusp condition (solid line) and without (dashed line) for the helium atom. With increase in basis set size, the correlation energy improves systematically and converges.\cite{note_he} In all basis sets, the $E_{corr}$ is lower when the cusp condition is enforced. In the smaller TZ2P basis set this effect is quite significant, resulting in a 5 mHa decrease in correlation energy at the FCI level. The effect diminishes to the sub-mHa level with larger basis sets (QZP and higher). The errors in the KS density (compared to the WF reference) behave similarly. With larger basis sets, the $L_1$ errors for the pairs of densities agree within $O(10^{-4})$ a.u. and converge to a common point (see Table \ref{tab:cuspbasis} in the supporting information). Differences remain in $L_1$ errors between the cusp-enforced vs. cusp-not-enforced densities at the TZ2P level, reflecting the smallness of this basis. For the larger basis set sizes, the cusp condition can readily be applied, giving an improved description of the density at the nucleus.

\begin{figure}[H]
    \includegraphics[width=\linewidth]{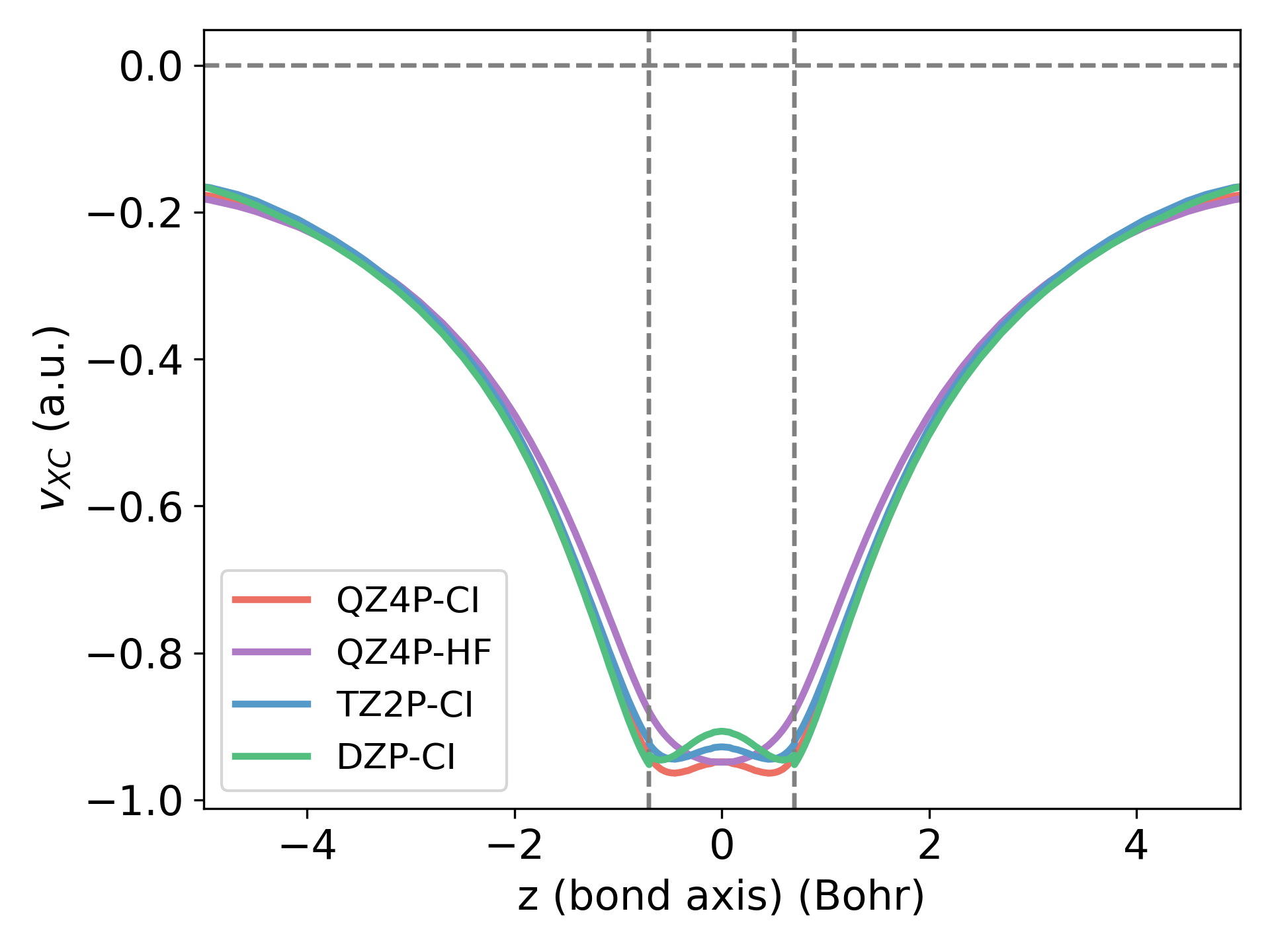}
    \caption{\label{fig:basis}The exchange correlation potential of $\mathrm{H_2}$(eq.) along the bond axis calculated using different basis sets and theoretical method. $v_{XC} = 0$ is shown as a horizontal grey dashed line.}
\end{figure}

Having established how the SlaterRKS procedure behaves with and without a cusp condition, the next item to examine is the effect of basis set size, this time in a diatomic molecule. Results for the $\mathrm{H_2}$ molecule (at its equilibrium geometry) with various basis sets are therefore shown in Fig. \ref{fig:basis}. The DZP basis set is too small to be meaningfully used alongside the cusp condition, but the others include the Kato cusp. Far from the nuclei where the density is low, the three different basis sets result in similar potentials. As more correlation is present near the bonding region, the larger basis sets (TZ, QZ) produce deeper wells in the exchange correlation potential, though the three basis sets give qualitatively similar structures consisting of double-well potentials. For comparison, the SlaterRKS potential from the HF wavefunction is also shown in Fig. \ref{fig:basis}. Notably, $v_{XC}$ from HF is missing the characteristic maximum in the middle of the bond, due to the lack of electron correlation.

\begin{figure}[H]
    \includegraphics[width=\linewidth]{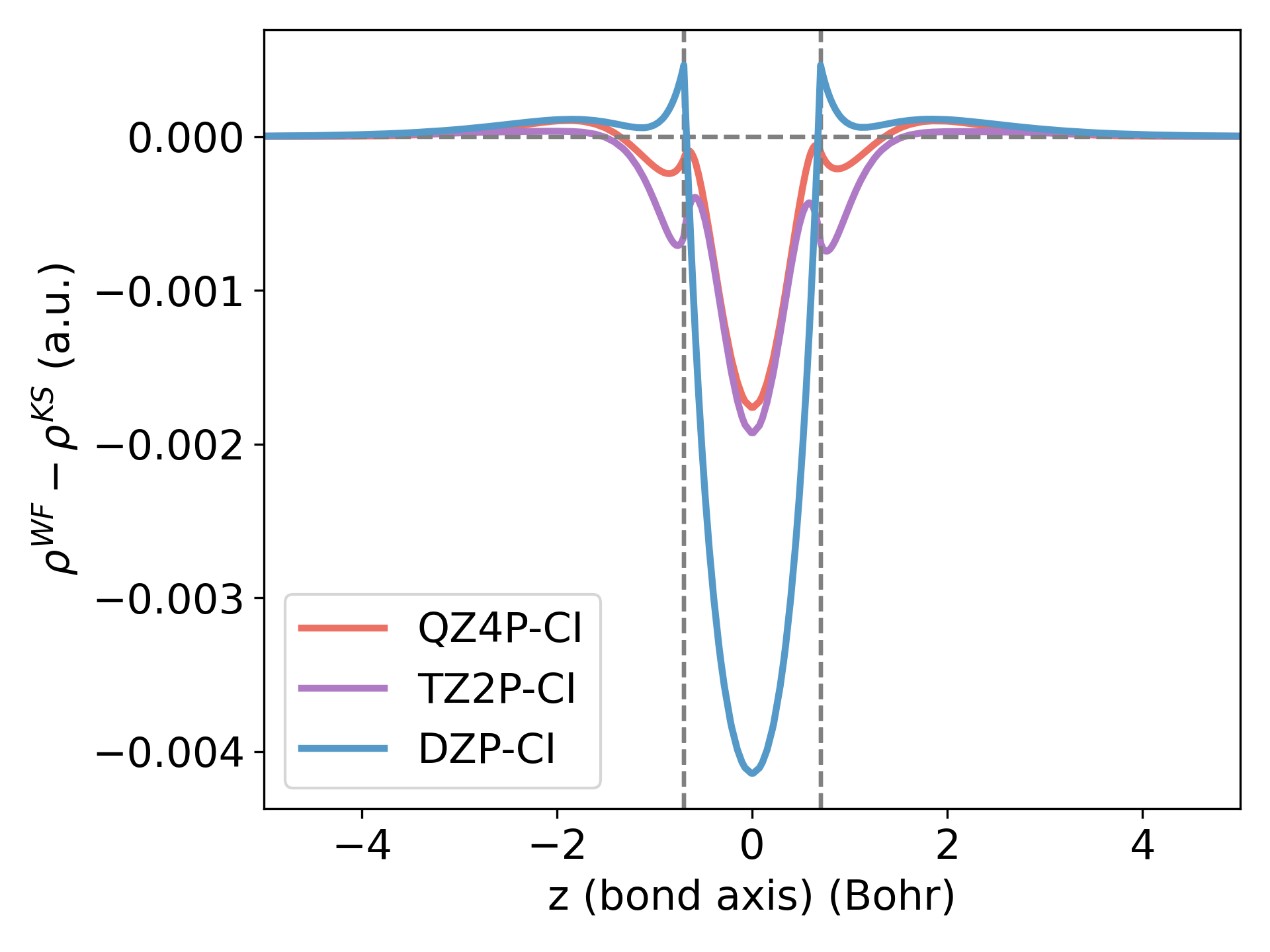}
    \caption{\label{fig:rhodiff}The difference in the FCI and RKS densities for $\mathrm{H_2}$(eq.) along the bond axis. $\rho^{WF}-\rho^{KS}=0$ is shown as a horizontal grey dashed line.}
\end{figure}

Fig. \ref{fig:rhodiff} shows the differences in density between FCI and SlaterRKS for the QZ4P, TZ2P and DZP basis sets. The corresponding $L_1$ and $L_2$ errors in the density for these calculations are given in the Supporting  Information. As reflected in the original RKS procedure in a Gaussian basis, larger basis sets do a better job at reproducing WF electron densities.\cite{cuevas-saavedra_kohnsham_2015,ryabinkin_reduction_2015} The Laplacian kinetic energy ($\tau_L$) can also be used in place of the positive-definite form, which is shown for the QZ4P basis set (see Supporting Information). The sharp features in the $\tau_L-v_{XC}$ are due to the divergence of the Laplacian near the nucleus, which arise due to the differences in WF and KS KE density via Eq. \ref{rks}. In all, because the larger (QZ4P) Slater basis set and positive-definite KE operator produced the best density (as well as potential), the remaining SlaterRKS results in this work were carried out using the QZ4P basis set, FCI RDM, and the positive-definite KE operator. 

\begin{figure}[H]
    \includegraphics[width=\linewidth]{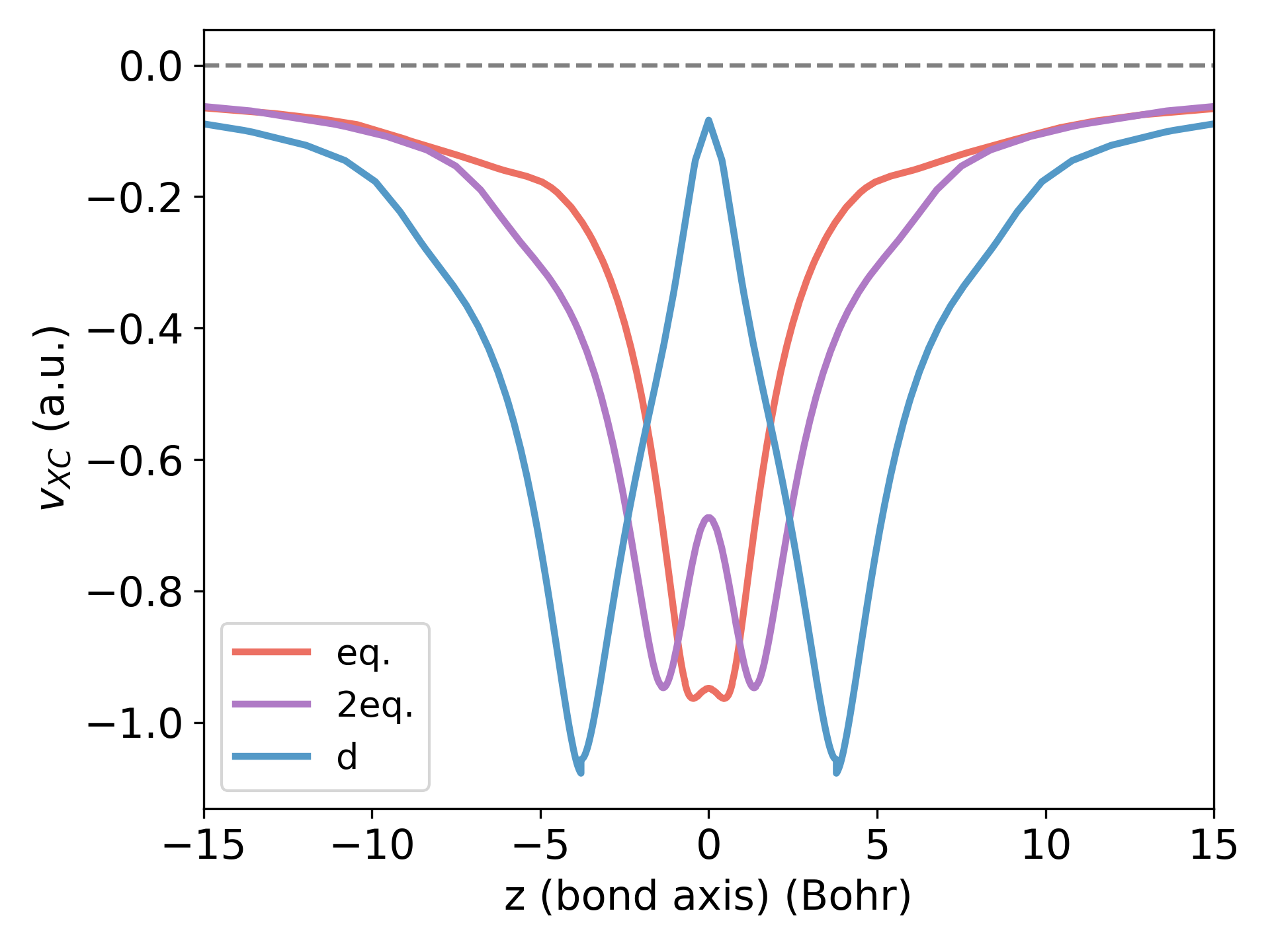}
    \caption{\label{fig:allh2}The exchange correlation potential of $\mathrm{H_2}$ along the bond axis with the hydrogen atoms spaced at 1.4010 a.u. (equilibrium distance) in red, 2.8020 a.u. (2 $\times$ equilibrium distance) in purple, and 7.58 a.u. (dissociation bond length) in blue. $v_{XC} = 0$ is shown as a horizontal grey dashed line.}
\end{figure}

\begin{table}[h]
\caption{\label{tab:table1} Normalized $L_1$ errors and $L_2$ errors of density from RKS with respect to CI densities in the QZ4P basis set}
\begin{ruledtabular}
\begin{tabular}{lcr}
System & $\Delta\rho_{L_1}/N_e$ & $\Delta\rho_{L_2}$\\
\hline
$\mathrm{H_2}$(eq.) & $4.06\times10^{-3}$ & $1.58\times10^{-3}$\\
$\mathrm{H_2}$(2eq.) & $1.37\times10^{-2}$ & $3.70\times10^{-3}$\\
$\mathrm{H_2}$(d) & $8.86\times10^{-3}$ & $1.49\times10^{-3}$ \\
LiH & $4.66\times10^{-3}$ & $4.68\times10^{-3}$ \\
$\mathrm{CH_2}$ & $4.91\times10^{-3}$ & $8.33\times10^{-3}$ \\
$\mathrm{H_2O}$ & $3.28\times10^{-3}$ & $1.88\times10^{-2}$
\end{tabular}
\end{ruledtabular}
\end{table}

To show the behavior of SlaterRKS as strong correlation is introduced to the reference WF, three geometries of the $\mathrm{H_2}$ molecule were examined. The $L_1$ errors of densities normalized by the number of electrons ($\Delta \rho_{L_1}/N_e$) are given in Table \ref{tab:table1} for the QZ4P basis set. Excellent agreement of the RKS and FCI densities in the case of the equilibrium geometry of $\mathrm{H_2}$ is demonstrated by the normalized $L_1$ error in the order of $10^{-3}$. The increase in the $L_1$ errors at the elongated bond lengths is expected as strong correlation increases since the FCI density has significantly noninteger orbital occupancies, which are challenging to reproduce with a single determinant in a finite basis set. The features of the $v_{XC}$s (shown along the H--H bond in Fig. \ref{fig:allh2}) are largely representative of the exact potentials obtained using the finite-element inverse calculation.\cite{kanungo_comparison_2021} For the strongly correlated cases of stretched $\mathrm{H_2}$ molecules, where most functionals fail, the SlaterRKS potentials are promising. The depth of the potential at the nucleus and the height of the maxima between the nuclei gradually increase as the bond length increases, signifying gradual depletion of electron density between the bond. 

\begin{figure}[H]
\includegraphics[width=\linewidth]{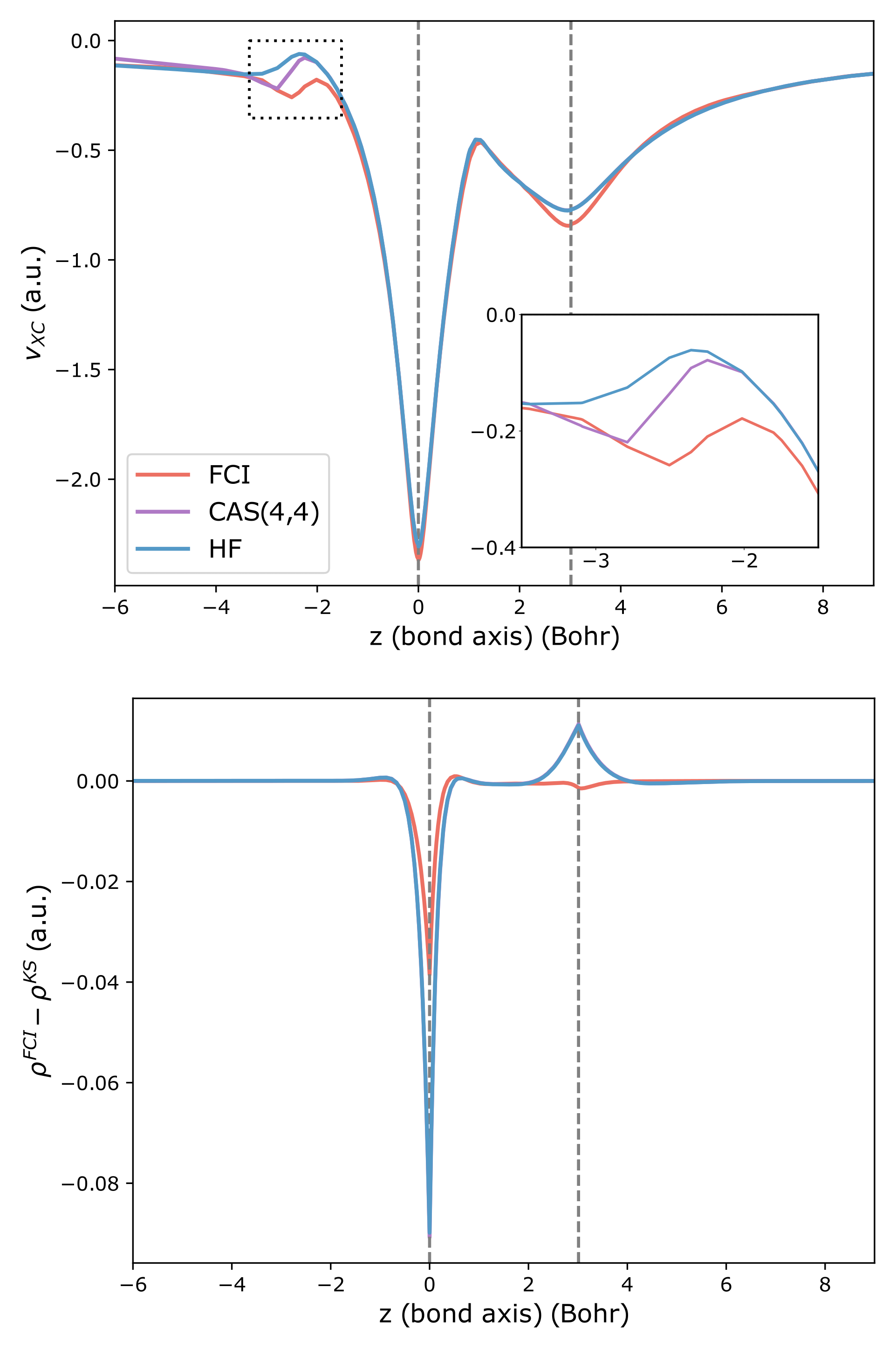}
\caption{\label{fig:LiHvxc} Top: The exchange correlation potential of LiH along the bond axis at the equilibrium separation of 3.01392 a.u. using RDMs from FCI, CAS(4,4) and HF. The vertical dashed lines represent the position of each nucleus. The inset shows a zoomed in part of the potential marked by dotted box. Bottom: The difference in densities $\rho^{FCI}-\rho^{KS}$ for each method.}
\end{figure}

Moving on to a diatomic with more electrons, SlaterRKS analysis for the minimum energy geometry of LiH is shown in Fig. \ref{fig:LiHvxc} (top). Various commonly used DFT XC potentials have a deeper well at Li, whereas the SlaterRKS $v_{XC}$ is shallower, closer to the accurate finite-element inverse calculation.\cite{kanungo_comparison_2021} A comparison with $v_{XC}$ profiles obtained from FCI, CASCI with a (4e,4o) active space and Hartree Fock (HF) reveals the role of fully correlated wavefunctions in RKS. The FCI has deeper wells at Li and H whereas both CASCI and HF have shallower wells and are almost indistinguishable (Fig. \ref{fig:LiHvxc}). To the left of the Li potential well, an intershell feature is present, typical of an accurate $v_{XC}$.\cite{kanungo_comparison_2021} The intershell feature gradually shifts away from Li in the sequence FCI to CAS to HF. The errors in the RKS densities constructed from various wavefunctions from the FCI density in the QZ4P basis set are shown in Fig. \ref{fig:LiHvxc} (bottom). The FCI RKS density has largest errors near the Li nucleus. This is likely the case because the QZ4P basis has only three core 1s orbitals, limiting the ability of SlaterRKS to resolve the density to higher accuracy. On the other hand, the CAS and HF RKS densities have larger errors at both Li and H due to lack of correlation. The corresponding $\Delta\rho_{L_1}/N_e$ of the CAS and HF RKS densities with respect to the FCI density are $1.32\times10^{-2}$ and $1.39\times10^{-2}$, respectively. Overall, the $L_1$ errors in the RKS density from the FCI reference are still relatively low, as reflected in Table \ref{tab:table1}. These might be improved further with the availability of larger Slater basis sets.


\begin{figure}[H]
\includegraphics[width=\linewidth]{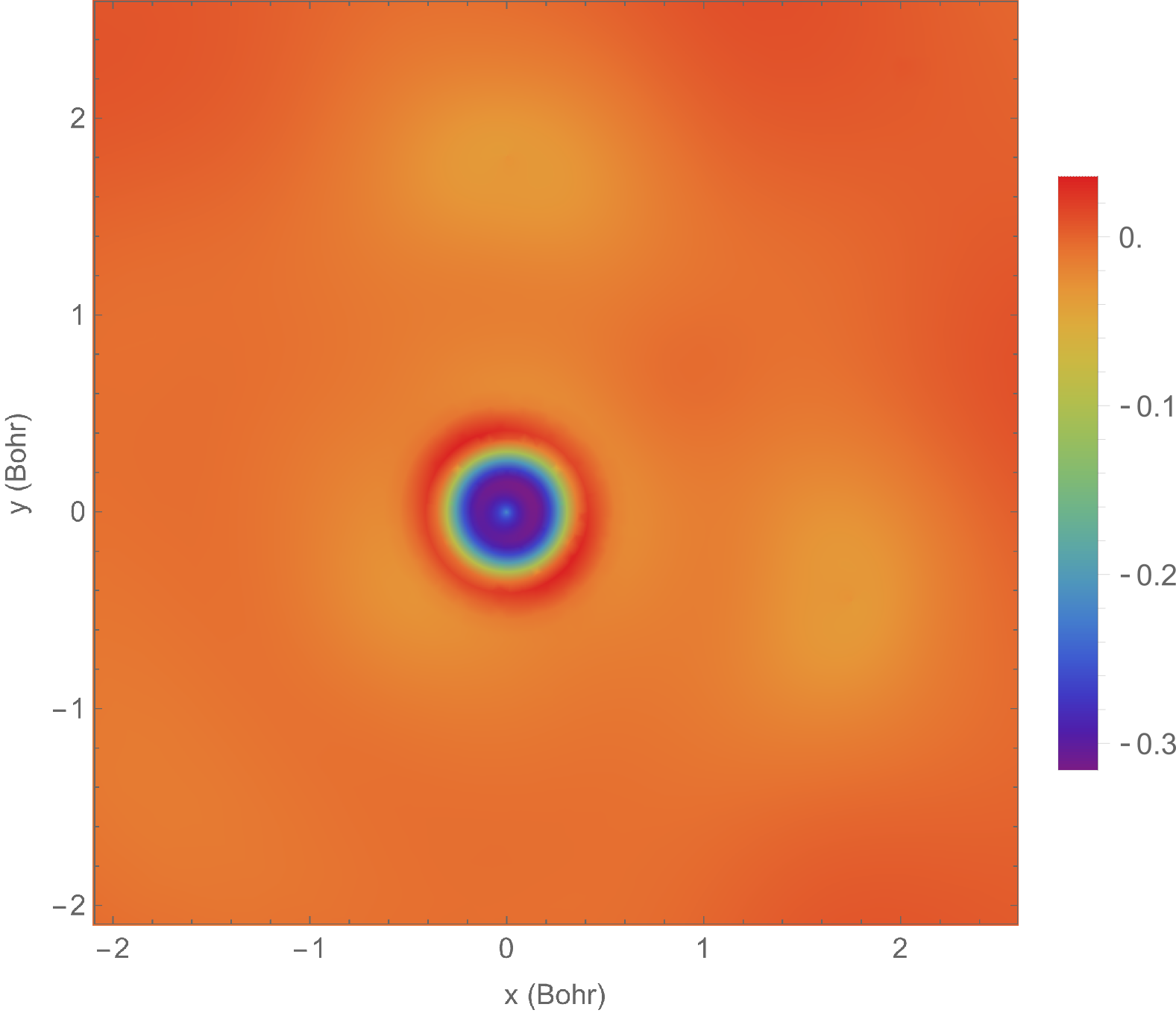}
\caption{\label{fig:H2O2dvxc} Comparison of exchange correlation potentials for $\mathrm{H_2O}$ in the plane of the molecule--difference between QZ4P and TZ2P $v_{XC}$ constructed from FCI RDM}
\end{figure}

The polyatomic $\mathrm{H_2O}$ molecule was next subjected to SlaterRKS analysis. The difference between QZ4P-FCI and TZ2P-FCI $v_{XC}$ is shown in Fig. \ref{fig:H2O2dvxc} in the molecular plane. The largest differences are near the O nuclei as the QZ4P basis set has a deeper well near $\mathbf{r}=0$. This difference gradually decreases as distance from the O nucleus increases until about 0.3-0.4 Bohr, after which a diffuse yellow band depicts the difference in the intershell feature in the two basis sets. The bonding region however is quite similar in both QZ and TZ with slightly higher differences appearing near the H nuclei.



\begin{figure*}
\includegraphics[width=\linewidth]{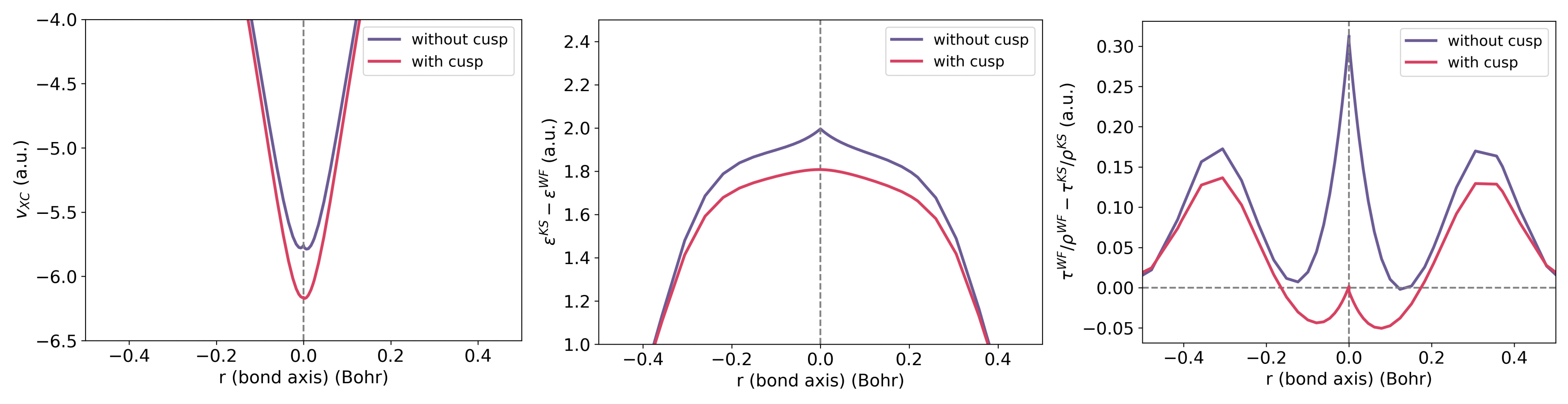}
\caption{\label{fig:H2O_cusp} The $v_{XC}$, $\epsilon^{KS}-\epsilon^{WF}$ and $\tau^{WF}/\rho^{WF}-\tau^{KS}/\rho^{KS}$ for $\mathrm{H_2O}$ zoomed in to $\mathbf{r}=0$ (O atom) position to demonstrate the effect of the cusp condition.}
\end{figure*}

The importance of having correct nuclear cusps was emphasized in the introduction, and an explicit enforcement of these cusps is present in the SlaterRKS algorithm. The $\mathrm{H_2O}$ molecule provides a good example to demonstrate the nature of SlaterRKS $v_{XC}$ with and without the cusp condition being enforced (Eq. \ref{kato}). Fig. \ref{fig:H2O_cusp} shows $v_{XC}$ in a small region around the oxygen nucleus. The peak that forms without the cusp condition is unexpected, and likely an artifact of using Slater orbitals that have steep slopes near the nucleus (see Eq. \ref{p}). Since Gaussian orbitals have zero slope at the nucleus, such behavior is less likely with finite-sized Gaussian basis sets. Fortunately, enforcement of the nuclear cusp condition dramatically remedies this situation, producing a more physical, single well potential near the oxygen nucleus. The middle and right side of Fig. \ref{fig:H2O_cusp} show that the better behavior in the cusp-enforced case stems from smoother contributions to $v_{XC}$ from $\epsilon$ and $\tau$, since the Slater potentials are monotonic (see SI). The contributions from $\epsilon$ and $\tau$ without the cusp condition have more features---which do not cancel out---and overall result in a significant, likely incorrect effect on $v_{XC}$.

\begin{figure}[H]
\includegraphics[width=\linewidth]{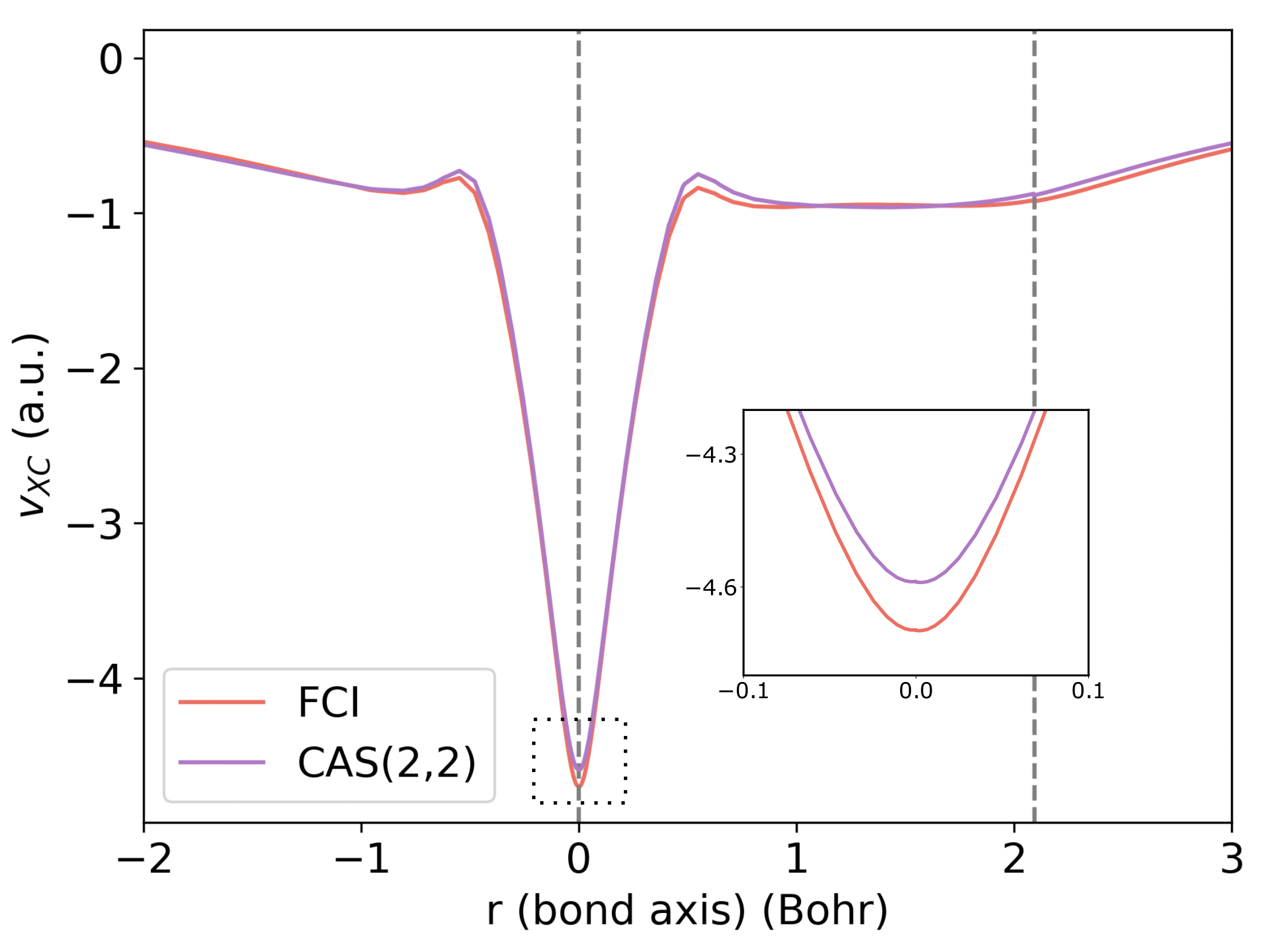}
\caption{\label{fig:CH2vxc} The exchange correlation potential of $\mathrm{CH_2}$ along one of the C--H bonds using RDMs from FCI and CAS(6,6). The C atom is at 0 a.u. and the H is at 2.0955 a.u. The vertical dashed lines represent the position of each nuclei. The inset shows a zoomed in part of the potential marked by dotted box.}
\end{figure}

Finally, SlaterRKS is used to examine a strongly correlated polyatomic. The lowest energy singlet state ($^1A_1$) of the $\mathrm{CH_2}$ (methylene) molecule has significant contributions from two electron configurations,
\begin{equation}
    (1a_1)^2(2a_1)^2(1b_2)^2(3a_1)^2
    \nonumber
\end{equation}
and,
\begin{equation}
    (1a_1)^2(2a_1)^2(1b_2)^2(1b_1)^2
    \nonumber 
\end{equation}
giving this state a multiconfigurational nature.\cite{sherrill_structures_1998,zimmerman_excited_2009,chien_excited_2018} Fig. \ref{fig:CH2vxc} shows the SlaterRKS XC potential along one of the C--H bonds. Using FCI the depth of the potential at the carbon nucelus is -4.7 a.u. which is in good agreement to previous reports.\cite{morrison_solution_1995,schipper_one_1998} Notably, the potential at both the nuclei is shallower when CASCI with a (2e,2o) active space is used. The intershell structure distinguishing the core and valence regions around the C atom is also present, as expected. This feature, however, is quantitatively different when the two methods are compared, with deviations on the order of 0.1 a.u. The potential at the H nucleus resembles the trailing valence region of the C atom. The nature of the $v_{XC}$ in the bonding region of the C--H bond of $\mathrm{CH_2}$ is relatively flat, which is a characteristic feature of covalent bonds.\cite{gritsenko_molecular_1996} 

The potential from Fig. \ref{fig:CH2vxc} results in an $L_1$ error in the electron density of $O(10^{-3})$, which is a satisfactory result given the complexity of the reference WF, which is from complete CI computations in the QZ4P basis set. The CI result includes population of  0.05958 electrons in the p orbital, a significant amount that is difficult to capture in the pure-state KS representation of the density. While this example has not been studied before using RKS theory, the SlaterRKS method appears up to the task of treating strong correlation in the challenging $\mathrm{CH_2}$ molecule.

\section{Conclusions}
This work found that RKS theory using a Slater basis set is a useful tool for examining exchange-correlation potentials corresponding to FCI wavefunctions. Features of the potentials match well with complete-basis-set results, compared to the finite-element inverse DFT results of Kanungo et al.\cite{kanungo_exact_2019,kanungo_comparison_2021} At the same time, however, quantitative accuracy of reproducing the FCI densities using moderately-sized Slater basis sets was also moderate, with $L_1$ errors on the order of $10^{-3}$ per electron. At the time of this work, Slater atomic orbital basis sets do not extend beyond quadruple zeta in quality, hindering progress towards more accurate densities via SlaterRKS. Future work to build larger, more complete Slater basis sets is likely to be instrumental in improving the accuracy of this method.

We anticipate that the SlaterRKS approach will be useful in providing facile analysis of strongly correlated molecules and their exchange-correlation potentials, using highly accurate reference wavefunctions (FCI) expressed in finite basis sets. The relative ease in convergence, physical behavior of the potentials, and correct asymptotics in the density are key advantages of SlaterRKS that merit further consideration of this method.

\section{Supplementary Material}

See the supplementary material for the geometries of the molecules, ionization energies, details of integration grid, description of convergence of cusp correction and the even-tempered basis set used for He, asymptotic behaviour of exchange correlation potential, basis set dependence of SlaterRKS method and effect of Laplacian kinetic energy densities. 

\section{Acknowledgements}

This project has been supported by the Department of Energy through the grant DE-SC0022241. The authors acknowledge the computing time on the Perlmutter Supercomputer from the National Energy Research Scientific Computing Center (NERSC) through allocation m4067. ST thanks support by the Eric and Wendy Schmidt AI in Science Postdoctoral Fellowship, a Schmidt Futures program.

\section{References}

\bibliography{citations}

\begin{thebibliography}{68}
\bibitem [{not()}]{note_he}%
  \BibitemOpen
  \href@noop {} {}\bibinfo {note} {The $E_{corr}$ converges at $\approx$ 37 mHa
  here, although the exact $E_{corr}$ is $\approx$ 42 mHa. This discrepancy is
  caused by the lack of higher angular momentum polarization functions in the
  basis sets. More details can be found in the Supporting
  Information}\BibitemShut {NoStop}%
\end{thebibliography}

\end{document}


\title{\centering Supplementary Material \\ \textit{for} \\ Exchange Correlation Potentials from Full Configuration Interaction in a Slater Orbital Basis} 

\author{Soumi Tribedi}
\affiliation{Department of Chemistry, University
of Michigan, Ann Arbor, Michigan 48109, United States}
\affiliation{Michigan Institute for Data Science, University of Michigan, Ann Arbor, Michigan 48109, United States}

\author{Duy-Khoi Dang}
\affiliation{Department of Chemistry, University
of Michigan, Ann Arbor, Michigan 48109, United States}

\author{Bikash Kanungo}
\affiliation{Department of Mechanical Engineering,
University of Michigan, Ann Arbor, Michigan 48109, United
States}

\author{Vikram Gavini}
\affiliation{Department of Mechanical Engineering,
University of Michigan, Ann Arbor, Michigan 48109, United
States}
\affiliation{Department of Materials Science and Engineering, University of Michigan, Ann Arbor, Michigan 48109, United
States}

\author{Paul M. Zimmerman}
\email{paulzim@umich.edu}
\affiliation{Department of Chemistry, University
of Michigan, Ann Arbor, Michigan 48109, United States}

\maketitle

\section{Coordinates}
The coordinates of all the molecules taken from the CCCBDB\cite{cccbdb} database are given below in Angstroms.
\vspace*{-\baselineskip}
\subsection{$\mathrm{H_2(eq.)}$}
\vspace*{-\baselineskip}
\begin{table}[H]
    \begin{tabular}{c c c c}
        H & 0.0000 & 0.0000 & 0.0000 \\
        H & 0.0000 & 0.0000 & 0.7414
    \end{tabular}
\end{table}

\subsection{$\mathrm{LiH}$}
\vspace*{-\baselineskip}
\begin{table}[H]
    \begin{tabular}{c c c c}
        Li & 0.0000 & 0.0000 & 0.0000 \\
        H & 0.0000 & 0.0000 & 1.5949 \\
    \end{tabular}
\end{table}

\subsection{$\mathrm{CH_2}$}
\vspace*{-\baselineskip}
\begin{table}[H]
    \begin{tabular}{c c c c}
        C & 0.0000 & 0.0000 & 0.0000 \\
        H & 0.0000 & 0.0000 & 1.1066 \\
        H & 0.0000 & 1.0809 & -0.2372
    \end{tabular}
\end{table}

\subsection{$\mathrm{H_2O}$}
\vspace*{-\baselineskip}
\begin{table}[H]
    \begin{tabular}{c c c c}
        O & 0.0000 & 0.0000 & 0.0000 \\
        H & 0.0000 & 0.0000 & 0.9578 \\
        H & 0.0000 & 0.9273 & -0.2395
    \end{tabular}
\end{table}

\section{Ionization Energies}
Experimental ionization energies were taken from CCCBDB\cite{cccbdb} database as references for the Kohn Sham eigenvalues.
\renewcommand{\thetable}{S\arabic{table}}
\begin{table}[H]
    \centering
    \caption{Experimental Ionization Energies}
    \begin{tabular}{c|c}
        Molecule & Ionization Energy (eV) \\
        \hline
        $\mathrm{H_2}$ & 15.42593 \\
        LiH & 7.9 \\
        $\mathrm{CH_2}$ & 10.396 \\
        $\mathrm{H_2O}$ & 12.621 \\
    \end{tabular}
    \label{tab:my_label}
\end{table}

\section{Integration Grid}
The dependence of the SlaterRKS potential on the angular grid size was tested for the $\mathrm{H_2}$ molecule (Figure \ref{fig:H2eq_grid}). The plot shows artifacts along the z axis (bonding axis of the molecule) at smaller angular grids. These artifacts disappear with larger grids, resulting in better behaved $v_{XC}$s. For all calculations in this manuscript, 50 radial and 5810 angular points were used.

\renewcommand{\thefigure}{S\arabic{figure}}
\begin{figure}[H]
\centering
\includegraphics[scale=0.53]{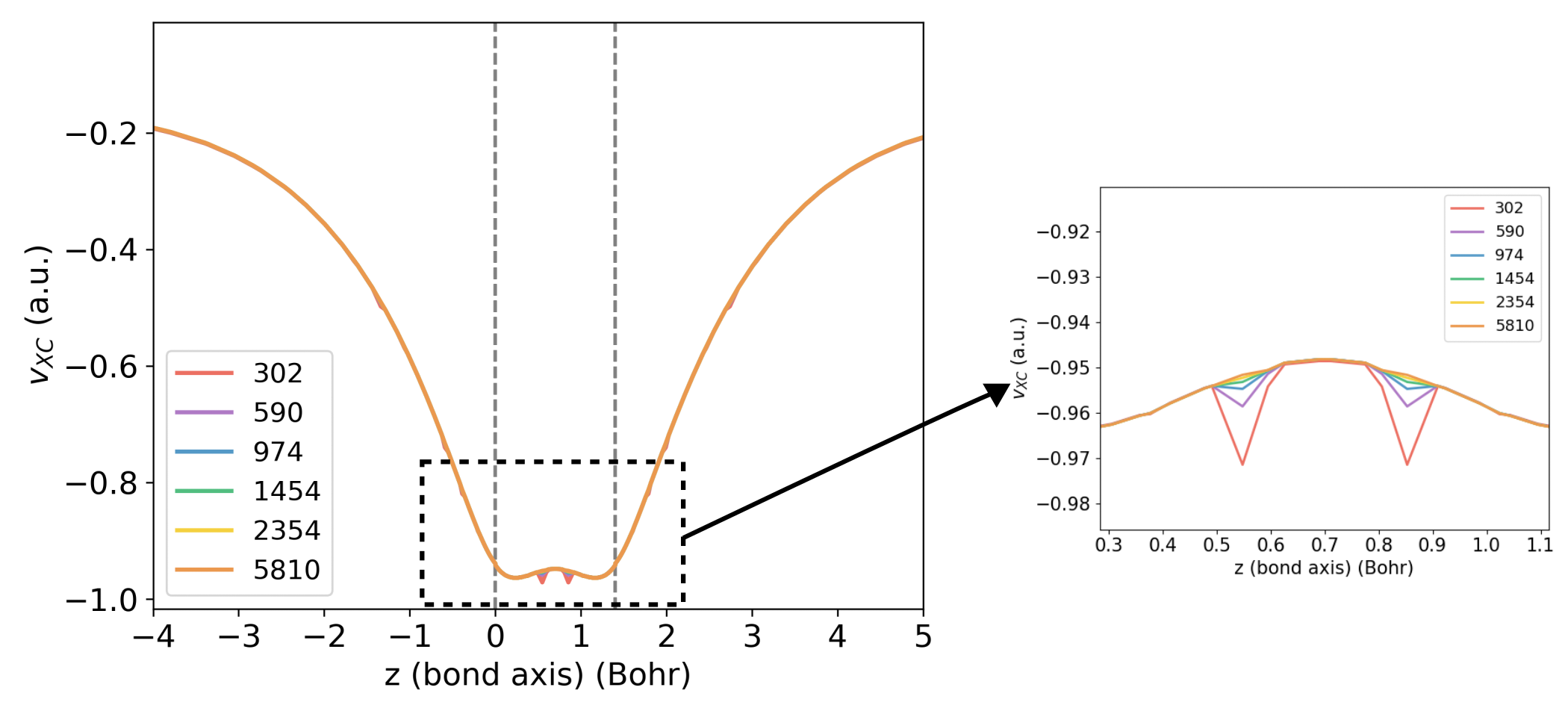}
\caption{\label{fig:H2eq_grid} The dependence of $v_{XC}$ on the angular grid points.}
\end{figure}


\section{Cusp Correction}
Even-tempered basis sets were created for the He atom in the 5Z and 6Z level in order to examine the cusp behaviour across a range of basis sets. Even-tempered basis set exponents for a given angular quantum number $l$ are determined by $\zeta_{lk} = \alpha_l (\beta_l)^k, k = 0,1,\dots,N-1$. The centroid exponent for s function is 1.6875, p 1.00, and d 2.00. The $\beta$ values are as follows: $\beta_s=1.795$, $\beta_p=1.800$ and $\beta_d=1.800$ as given in Baerends and coworkers.\cite{chong_even-tempered_2004} It is important to note that the centroid exponents and $\beta$ values for He were only available up to the d function, therefore basis sets containing higher angular momentum functions which are necessary to account for the entire correlation could not be constructed and hence the correlation energy at even the biggest basis set is short of the exact CBS limit value by about 5 mHa (Fig. 2)

\subsection*{5Z6P}
\vspace{-1cm}
\begin{table}[H]
    \begin{tabular}{l l}
        1S & 5.4372 \\
        1S & 3.0291 \\
        1S & 1.6875 \\
        1S & 0.9401 \\
        1S & 0.5237 \\
        2P & 1.8000 \\
        2P & 1.0000 \\
        2P & 0.5556 \\
        3D & 3.6000 \\
        3D & 2.0000 \\
        3D & 1.1111 
    \end{tabular}
\end{table}

\subsection*{6Z6P}
\vspace*{-\baselineskip}
\begin{table}[H]
    \begin{tabular}{l l}
        1S & 6.9836 \\
        1S & 3.8906 \\
        1S & 2.1675 \\
        1S & 1.2075 \\
        1S & 0.6727 \\
        1S & 0.3748 \\
        2P & 1.8000 \\
        2P & 1.0000 \\
        2P & 0.5556 \\
        3D & 3.6000 \\
        3D & 2.0000 \\
        3D & 1.1111 
    \end{tabular}
\end{table}

The cusp correction is implemented according to Handy's formulation\cite{handy_molecular_2004} for 1s and 2s Slater functions. Below in Tables \ref{tab:cusptz}-\ref{tab:cusp6z}, the left hand and right hand side of Kato's cusp condition (Eq. \ref{cuspcondition}) is evaluated at radial grid points close to the nucleus of a He atom using TZ2P, QZ4P, 5Z6P and 6Z6P basis sets. The difference $-2Z\rho(\mathbf{r}) - \frac{\partial \rho(\mathbf{r})}{\partial \mathbf{r}}$ is evaluated at these grid points and then the fractional error $\big(2Z\rho(\mathbf{r}) + \frac{\partial \rho(\mathbf{r})}{\partial \mathbf{r}}\big)/2Z\rho(\mathbf{r})$ is examined. 
The results are tabulated for the cusp corrected and uncorrected calculations. For a cusp corrected calculation, Kato's theorem is valid for up to 3-4 decimal points, whereas when the cusp condition has not been enforced by Handy's formulation, they disagree at the second decimal point.

\renewcommand{\theequation}{S\arabic{equation}}
\begin{equation}
    \frac{\partial \rho(\mathbf{r})}{\partial \mathbf{r}}\bigg|_{\mathbf{r}=R_B} = -2Z_B\rho(\mathbf{r})|_{\mathbf{r}=R_B}
    \label{cuspcondition}
\end{equation}

\begin{table}[H]
    \centering
    \caption{Comparison of cusp condition evaluated at radial points for He atom in the TZ2P basis set. \textbf{P}$=-2Z\rho(\mathbf{r})$; \textbf{Q}$=\frac{\partial \rho(\mathbf{r})}{\partial \mathbf{r}}$; \textbf{R}$=\frac{2Z\rho(\mathbf{r}) + \partial \rho(\mathbf{r})/\partial \mathbf{r}}{2Z\rho(\mathbf{r})}$}
    \begin{ruledtabular}
    \begin{tabular}{c|c|c|c|c|c|c|c}
         \multicolumn{2}{c|}{Grid points} & \multicolumn{3}{c|}{Without cusp} & \multicolumn{3}{c}{With cusp} \\
        \hline
        $\mathbf{r}$ & $\mathbf{r}+1$ & \textbf{P} & \textbf{Q} & \textbf{R} & \textbf{P} & \textbf{Q} & \textbf{R}\\
        \hline
         0.00000500 & 0.00013500 & -14.294355 & -14.224191 & 0.00491 & -14.392900 & -14.391798 & 0.00008\\
         0.00013500 & 0.00062504 & -14.276732 & -14.207437 & 0.00485 & -14.375070 & -14.374317 & 0.00005\\
         0.00062504 & 0.00171529 & -14.231932 & -14.159590 & 0.00508 & -14.329745 & -14.325617 & 0.00029\\
         0.00171529 & 0.00364633 & -14.146719 & -14.069657 & 0.00545 & -14.243535 & -14.234085 & 0.00066\\
         0.00364633 & 0.00665943 & -14.008475 & -13.923594 & 0.00606 & -14.103680 & -14.085560 & 0.00128 \\
         \hline
         \multicolumn{2}{c|}{Total CI energy:} &
         \multicolumn{3}{c|}{-2.88075985} &
         \multicolumn{3}{c}{-2.87539229}
    \end{tabular}
    \end{ruledtabular}
    \label{tab:cusptz}
\end{table}

\begin{table}[H]
    \centering
    \caption{Comparison of cusp condition evaluated at radial points for He atom in the QZ4P basis set. \textbf{P}$=-2Z\rho(\mathbf{r})$; \textbf{Q}$=\frac{\partial \rho(\mathbf{r})}{\partial \mathbf{r}}$; \textbf{R}$=\frac{2Z\rho(\mathbf{r}) + \partial \rho(\mathbf{r})/\partial \mathbf{r}}{2Z\rho(\mathbf{r})}$}
    \begin{ruledtabular}
    \begin{tabular}{c|c|c|c|c|c|c|c}
         \multicolumn{2}{c|}{Grid points} & \multicolumn{3}{c|}{Without cusp} & \multicolumn{3}{c}{With cusp} \\
        \hline
        $\mathbf{r}$ & $\mathbf{r}+1$ & \textbf{P} & \textbf{Q} & \textbf{R} & \textbf{P} & \textbf{Q} & \textbf{R}\\
        \hline
         0.00000500 & 0.00013500 & -14.418209 & -14.581808 & -0.01135 & -14.344258 & -14.345716 & -0.00010 \\
         0.00013500 & 0.00062504 & -14.400145 & -14.563085 & -0.01132 & -14.326488 & -14.325626 & 0.00006 \\
         0.00062504 & 0.00171529 & -14.354227 & -14.512564 & -0.01103 & -14.281316 & -14.277287 & 0.00028 \\
         0.00171529 & 0.00364633 & -14.266901 & -14.417337 & -0.01054 & -14.195396 & -14.186107 & 0.00065 \\
         0.00364633 & 0.00665943 & -14.125267 & -14.263328 & -0.00977 & -14.056011 & -14.038234 & 0.00126 \\
         \hline
         \multicolumn{2}{c|}{Total CI energy:} &
         \multicolumn{3}{c|}{-2.89333122} &
         \multicolumn{3}{c}{-2.89307691}
    \end{tabular}
    \end{ruledtabular}
    \label{tab:cuspqz}
\end{table}

\begin{table}[H]
    \centering
    \caption{Comparison of cusp condition evaluated at radial points for He atom in the 5Z6P basis set. \textbf{P}$=-2Z\rho(\mathbf{r})$; \textbf{Q}$=\frac{\partial \rho(\mathbf{r})}{\partial \mathbf{r}}$; \textbf{R}$=\frac{2Z\rho(\mathbf{r}) + \partial \rho(\mathbf{r})/\partial \mathbf{r}}{2Z\rho(\mathbf{r})}$}
    \begin{ruledtabular}
    \begin{tabular}{c|c|c|c|c|c|c|c}
         \multicolumn{2}{c|}{Grid points} & \multicolumn{3}{c|}{Without cusp} & \multicolumn{3}{c}{With cusp} \\
        \hline
        $\mathbf{r}$ & $\mathbf{r}+1$ & \textbf{P} & \textbf{Q} & \textbf{R} & \textbf{P} & \textbf{Q} & \textbf{R}\\
        \hline
        0.00000500 & 0.00013500 & -14.509079 & -14.943793 & -0.02996 & -14.266931 & -14.266032 & 0.00006\\
        0.00013500 & 0.00062504 & -14.490567 & -14.923506 & -0.02988 & -14.249258 & -14.247479 & 0.00012\\
        0.00062504 & 0.00171529 & -14.443516 & -14.870292 & -0.02955 & -14.204332 & -14.199726 & 0.00032\\
        0.00171529 & 0.00364633 & -14.354055 & -14.768267 & -0.02886 & -14.118885 & -14.107388 & 0.00081\\
        0.00364633 & 0.00665943 & -14.209014 & -14.603803 & -0.02778 & -13.980284 & -13.958629 & 0.00155 \\
        \hline
         \multicolumn{2}{c|}{Total CI energy:} &
         \multicolumn{3}{c|}{-2.89867811} &
         \multicolumn{3}{c}{-2.89830824}
    \end{tabular}
    \end{ruledtabular}
    \label{tab:cusp5z}
\end{table}

\begin{table}[H]
    \centering
    \caption{Comparison of cusp condition evaluated at radial points for He atom in the 6Z6P basis set. \textbf{P}$=-2Z\rho(\mathbf{r})$; \textbf{Q}$=\frac{\partial \rho(\mathbf{r})}{\partial \mathbf{r}}$; \textbf{R}$=\frac{2Z\rho(\mathbf{r}) + \partial \rho(\mathbf{r})/\partial \mathbf{r}}{2Z\rho(\mathbf{r})}$}
    \begin{ruledtabular}
    \begin{tabular}{c|c|c|c|c|c|c|c}
         \multicolumn{2}{c|}{Grid points} & \multicolumn{3}{c|}{Without cusp} & \multicolumn{3}{c}{With cusp} \\
        \hline
        $\mathbf{r}$ & $\mathbf{r}+1$ & \textbf{P} & \textbf{Q} & \textbf{R} & \textbf{P} & \textbf{Q} & \textbf{R}\\
        \hline
        0.00000500 & 0.00013500 & -14.444248 & -14.731427 & -0.01988 & -14.330941 & -14.334082 & -0.00022\\
        0.00013500 & 0.00062504 & -14.426004 & -14.707119 & -0.01949 & -14.313188 & -14.311708 & 0.00010\\
        0.00062504 & 0.00171529 & -14.379632 & -14.656072 & -0.01922 & -14.268059 & -14.263782 & 0.00030\\
        0.00171529 & 0.00364633 & -14.291453 & -14.557444 & -0.01861 & -14.182222 & -14.172362 & 0.00070\\
        0.00364633 & 0.00665943 & -14.148466 & -14.397860 & -0.01763 & -14.042975 & -14.024024 & 0.00135 \\
        \hline
         \multicolumn{2}{c|}{Total CI energy:} &
         \multicolumn{3}{c|}{-2.89889649} &
         \multicolumn{3}{c}{-2.89877522}
    \end{tabular}
    \end{ruledtabular}
    \label{tab:cusp6z}
\end{table}

\begin{table}[H]
\caption{Normalized $L_1$ errors and $L_2$ errors of density from RKS in comparison to CI densities for He molecule with and without cusp correction}
\label{tab:cuspbasis}
\begin{ruledtabular}
\begin{tabular}{llll}
Basis Set & Cusp Correction & $\Delta\rho_{L1}$  & $\Delta\rho_{L2}$\\
\hline
TZ2P & Yes & $1.25\times10^{-2}$ & $2.64\times10^{-3}$\\
TZ2P & No & $8.32\times10^{-3}$ & $3.13\times10^{-3}$ \\
QZ4P & Yes & $9.12\times10^{-3}$ & $2.39\times10^{-3}$ \\
QZ4P & No & $9.16\times10^{-3}$ & $2.26\times10^{-3}$ \\
5Z6P & Yes & $6.75\times10^{-3}$ & $1.98\times10^{-3}$ \\
5Z6P & No & $7.04\times10^{-3}$ & $2.09\times10^{-3}$ \\
6Z6P & Yes & $6.76\times10^{-3}$ & $2.00\times10^{-3}$ \\
6Z6P & No & $6.80\times10^{-3}$ & $2.10\times10^{-3}$ \\
\end{tabular}
\end{ruledtabular}
\end{table}

\section{Asymptotic behaviour}
$v_{XC}$ and $v_{XC,Slater}^{WF}$ are expected to decay asymptotically as $-1/R$. This is shown for the $\mathrm{H_2}$ molecule in the QZ4P basis set.

\begin{figure}[H]
\centering
\includegraphics[scale=0.56]{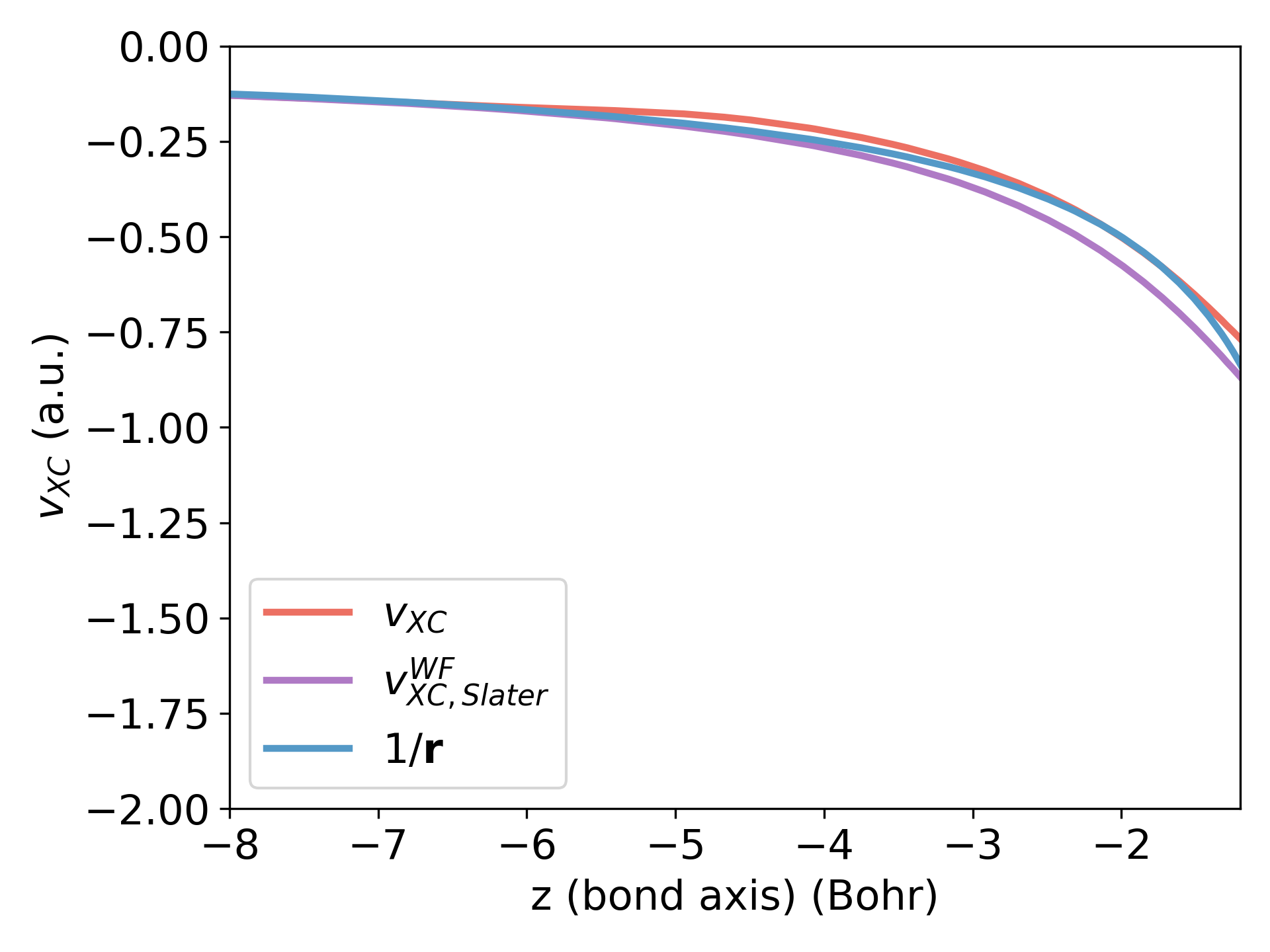}
\caption{\label{fig:H21R} Comparison of $v_{XC}$ and $v_S^{WF}$ with $1/R$ for $\mathrm{H_2}$(eq.) at larger distances ($\mathbf{r}$) from the nuclei.}
\end{figure}

\section{Basis set dependence}
The basis set dependence on the SlaterRKS method is demonstrated in Table \ref{tab:basis_tab} using the $\mathrm{H_2}$ molecule at its equilibrium geometry. The comparison between FCI and HF RDM is also made for the QZ4P basis set. SlaterRKS method recovers the HF density more effectively than the FCI density.

\begin{table}[h!]
\caption{\label{tab:basis_tab} Normalized $L_1$ errors and $L_2$ errors of density from RKS in comparison to wavefunction densities for $\mathrm{H_2}$(eq.) molecule.}
\begin{ruledtabular}
\begin{tabular}{llll}
Basis Set & Method & $\Delta\rho_{L1}/N_e$ & $\Delta\rho_{L2}$\\
\hline
DZP(no cusp) & FCI & $9.71\times10^{-3}$ & $3.77\times10^{-3}$ \\
TZ2P & FCI & $6.81\times10^{-3}$ & $2.25\times10^{-3}$\\
QZ4P & FCI & $4.06\times10^{-3}$ & $1.58\times10^{-3}$\\
QZ4P & HF & $3.95\times10^{-4}$ & $1.00\times10^{-4}$
\end{tabular}
\end{ruledtabular}
\end{table}

\section{Kinetic energy density}
\begin{figure}[H]
\centering
\includegraphics[scale=0.53]{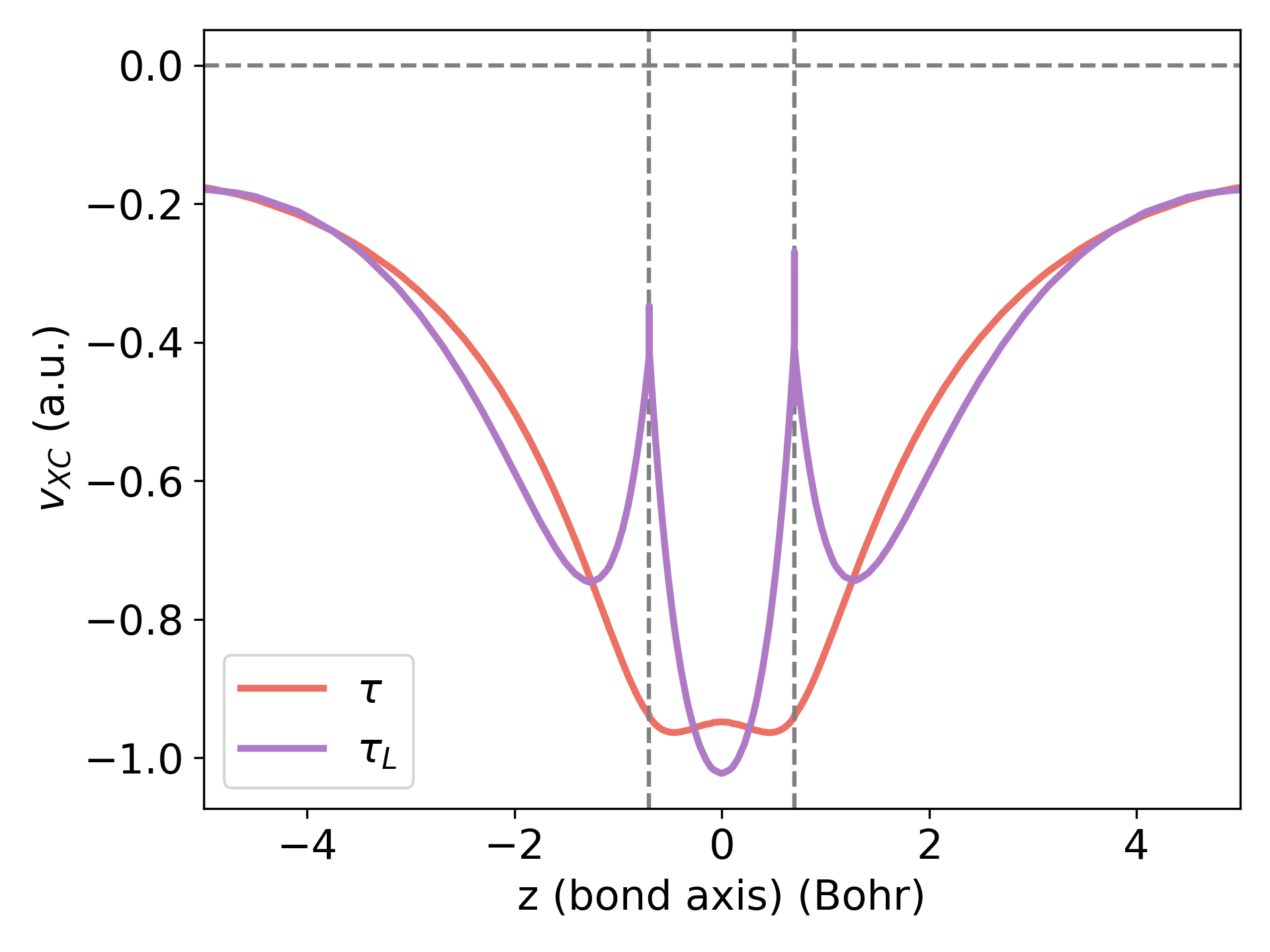}
\caption{\label{fig:H2eq_tl} The comparison of $v_{XC}$ using the Laplacian KE ($\tau_L$) and positive-definite KE ($\tau$) calculated using FCI at the QZ4P basis set.}
\end{figure}

\begin{figure}[H]
    \centering
    \includegraphics[scale=0.53]{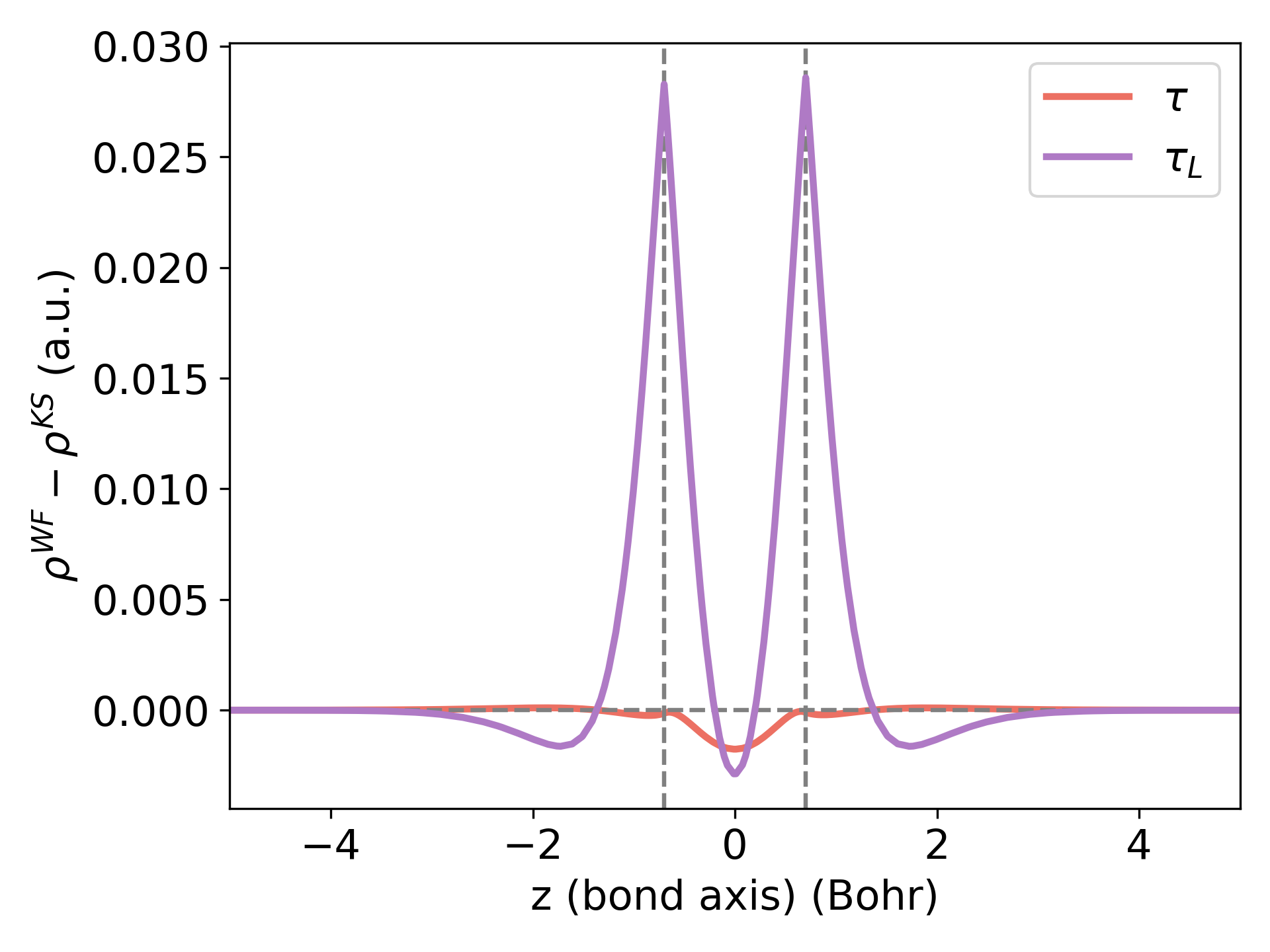}
    \caption{\label{fig:tl_rhodiff} The comparison of the difference in the RKS density from the FCI density calculated at the QZ4P basis set.}
\end{figure}

The $v_{XC}$ using the Laplacian KE ($\tau_L$) with QZ4P basis FCI RDM is shown in comparison to the positive-definite KE ($\tau$) in Figure \ref{fig:H2eq_tl}. With $\tau_L$ the potentials are inaccurate at the nuclear position and the bonding region, whereas it is qualitatively correct in other regions. The respective errors in densities are shown in Figure \ref{fig:tl_rhodiff}, where consequently the highest error in density with the Laplacian KE density is at the nuclear positions.

\bibliography{citations}